# AI Sustainability in Practice

## Part One: Foundations for Sustainable AI Projects

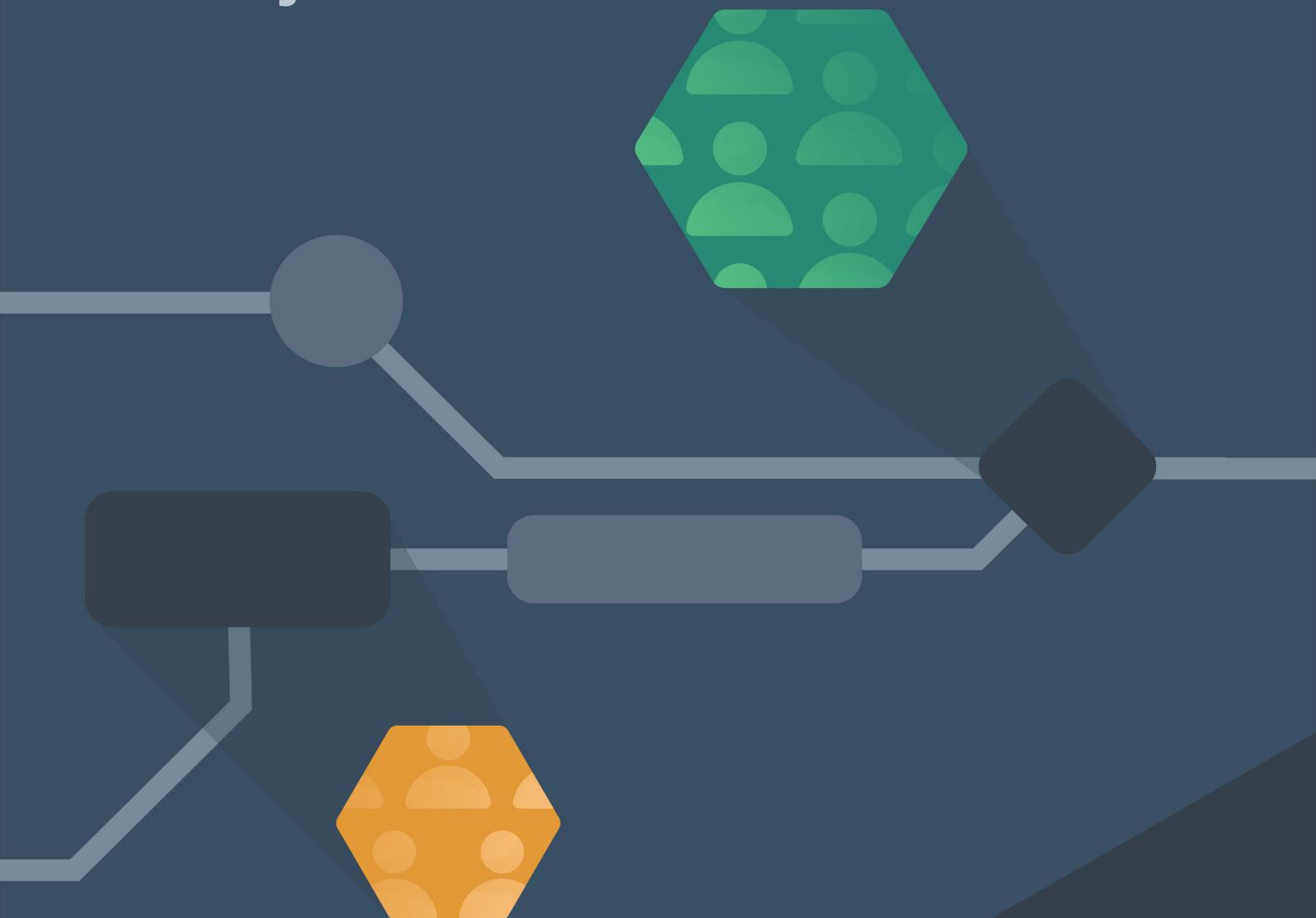

**For Facilitators**

This workbook is annotated to support facilitators in delivering the accompanying activities.

The Alan Turing Institute

# Acknowledgements

This workbook was written by David Leslie, Cami Rincón, Morgan Briggs, Antonella Perini, Smera Jayadeva, Ann Borda, SJ Bennett, Christopher Burr, Mhairi Aitken, Michael Katell, Claudia Fischer, Janis Wong, and Ismael Kherroubi Garcia.

The creation of this workbook would not have been possible without the support and efforts of various partners and collaborators. As ever, all members of our brilliant team of researchers in the Ethics Theme of the Public Policy Programme at The Alan Turing Institute have been crucial and inimitable supports of this project from its inception several years ago, as have our Public Policy Programme Co-Directors, Helen Margetts and Cosmina Dorobantu. We are deeply thankful to Conor Rigby, who led the design of this workbook and provided extraordinary feedback across its iterations. We also want to acknowledge Johnny Lighthands, who created various illustrations for this document, and Alex Krook and John Gilbert, whose input and insights helped get the workbook over the finish line. Special thanks must be given to the Ministry of Justice for helping us test the activities and review the content included in this workbook. Lastly, we want to thank Semeli Hadjiloizou (The Alan Turing Institute) for her meticulous peer review and timely feedback, which greatly enriched this document.

This work was supported by Wave 1 of The UKRI Strategic Priorities Fund under the EPSRC Grant EP/W006022/1, particularly the Public Policy Programme theme within that grant & The Alan Turing Institute; Towards Turing 2.0 under the EPSRC Grant EP/W037211/1 & The Alan Turing Institute; and the Ecosystem Leadership Award under the EPSRC Grant EP/X03870X/1 & The Alan Turing Institute.

Cite this work as: Leslie, D., Rincón, C., Briggs, M., Perini, A., Jayadeva, S., Borda, A., Bennett, SJ. Burr, C., Aitken, M., Katell, M., Fischer, C., Wong, J., and Kherroubi Garcia, I. (2023). *AI Sustainability in Practice. Part One: Foundations for Sustainable AI Projects.* The Alan Turing Institute.

# Contents

# About the AI Ethics and Governance in Practice Workbook Series

## Who We Are

The Public Policy Programme at The Alan Turing Institute was set up in May 2018 with the aim of developing research, tools, and techniques that help governments innovate with data-intensive technologies and improve the quality of people's lives. We work alongside policymakers to explore how data science and artificial intelligence can inform public policy and improve the provision of public services. We believe that governments can reap the benefits of these technologies only if they make considerations of ethics and safety a first priority.

## Origins of the Workbook Series

In 2019, The Alan Turing Institute's Public Policy Programme, in collaboration with the UK's Office for Artificial Intelligence and the Government Digital Service, published the UK Government's official Public Sector Guidance on AI Ethics and Safety. This document provides end-to-end guidance on how to apply principles of AI ethics and safety to the design, development, and implementation of algorithmic systems in the public sector. It provides a governance framework designed to assist AI project teams in ensuring that the AI technologies they build, procure, or use are ethical, safe, and responsible.

In 2021, the UK's National AI Strategy recommended as a 'key action' the update and expansion of this original guidance. From 2021 to 2023, with the support of funding from the Office for AI and the Engineering and Physical Sciences Research Council as well as with the assistance of several public sector bodies, we undertook this updating and expansion. The result is the AI Ethics and Governance in Practice Programme, a bespoke series of eight workbooks and a forthcoming digital platform designed to equip the public sector with tools, training, and support for adopting what we call a Process-Based Governance (PBG) Framework to carry out projects in line with state-of-the-art practices in responsible and trustworthy AI innovation.

# About the Workbooks

The AI Ethics and Governance in Practice Programme curriculum is composed of a series of eight workbooks. Each of the workbooks in the series covers how to implement a key component of the PBG Framework. These include Sustainability, Technical Safety, Accountability, Fairness, Explainability, and Data Stewardship. Each of the workbooks also focuses on a specific domain, so that case studies can be used to promote ethical reflection and animate the Key Concepts.

## Programme Curriculum: AI Ethics and Governance in Practice Workbook Series

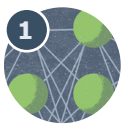 **AI Ethics and Governance in Practice: An Introduction**
*Multiple Domains*

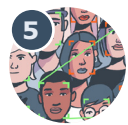 **Responsible Data Stewardship in Practice**
*AI in Policing and Criminal Justice*

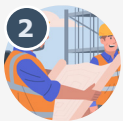 **AI Sustainability in Practice Part One**
*AI in Urban Planning*

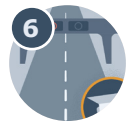 **AI Safety in Practice**
*AI in Transport*

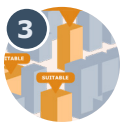 **AI Sustainability in Practice Part Two**
*AI in Urban Planning*

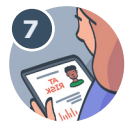 **AI Transparency and Explainability in Practice**
*AI in Social Care*

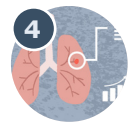 **AI Fairness in Practice**
*AI in Healthcare*

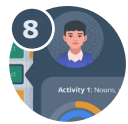 **AI Accountability in Practice**
*AI in Education*

Taken together, the workbooks are intended to provide public sector bodies with the skills required for putting AI ethics and governance principles into practice through the full implementation of the guidance. To this end, they contain activities with instructions for either facilitating or participating in capacity-building workshops.

Please note, these workbooks are living documents that will evolve and improve with input from users, affected stakeholders, and interested parties. We need your participation. Please share feedback with us at policy@turing.ac.uk.

## Programme Roadmap

The graphic below visualises this workbook in context alongside key frameworks, values and principles discussed within this programme. For more information on how these elements build upon one another, refer to AI Ethics and Governance in Practice: An Introduction.

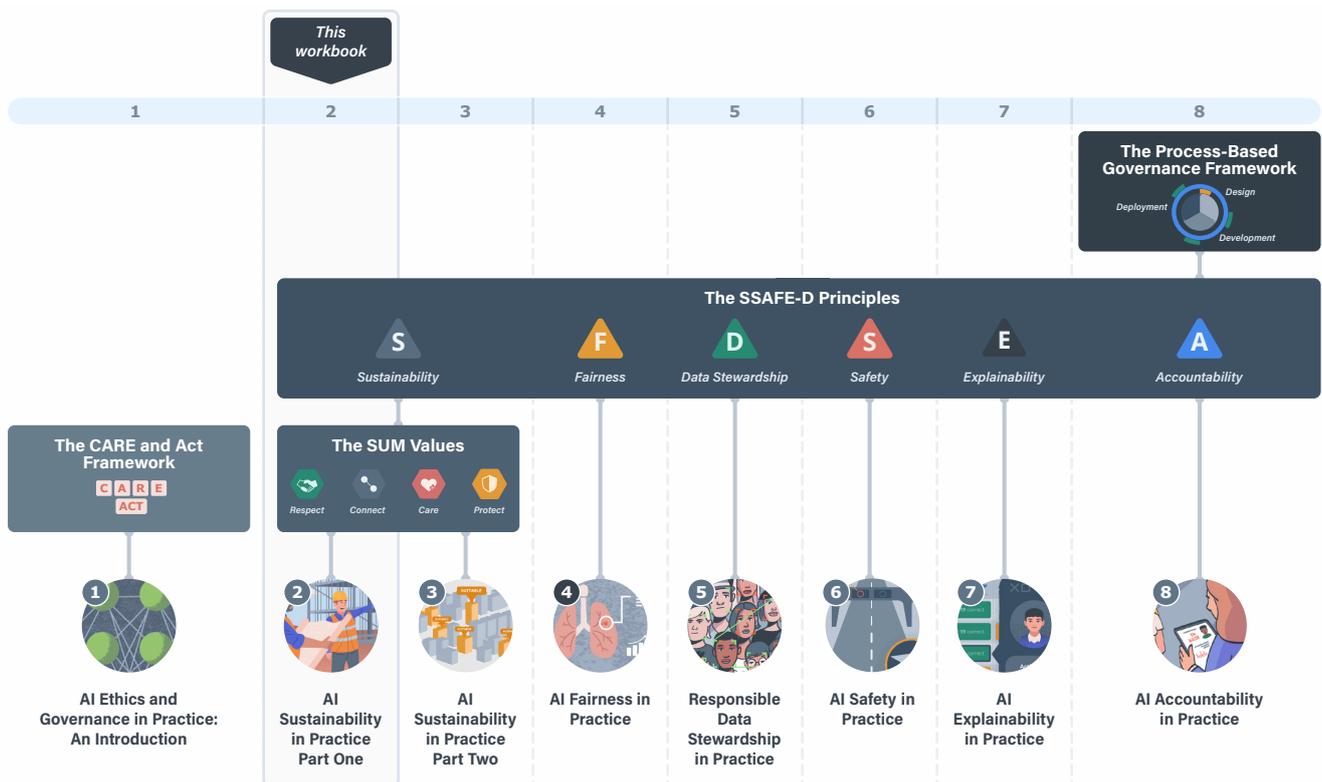

# Intended Audience

The workbooks are primarily aimed at civil servants engaging in the AI Ethics and Governance in Practice Programme as either AI Ethics Champions delivering the curriculum within their organisations by facilitating peer-learning workshops, or participants completing the programme by attending workshops. Anyone interested in learning about AI ethics, however, can make use of the programme curriculum, the workbooks, and resources provided. These have been designed to serve as stand-alone, open access resources. Find out more at turing.ac.uk/ai-ethics-governance.

There are two versions of each workbook:

- **Annotated workbooks** (such as this document) are intended for facilitators. These contain guidance and resources for preparing and facilitating training workshops.

- **Non-annotated workbooks** are intended for workshop participants to engage with in preparation for, and during, workshops.

# Introduction to This Workbook

This workbook is part one of two workbooks: Foundations for Sustainable AI Projects and Sustainability Throughout the AI Workflow. Both workbooks are intended to help faciliate the delivery of a two-part workshop on the concepts of SUM Values and Sustainability.

### AI Sustainability in Practice Part One: Foundations for Sustainable AI Projects

This workbook introduces the SUM Values and the principle of Sustainability, which help AI project teams to assess the potential societal impacts and ethical permissibility of their projects. This workbook is divided into two sections, key Concepts and Activities:

<div align="center">

**Key Concepts Section**

</div>

This section discusses frameworks for establishing the foundations for sustainable AI projects:

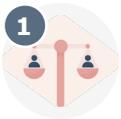

Introduction to Sustainability: The SUM Values

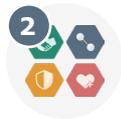

The SUM Values in Focus: Respect, Connect, Care, and Protect

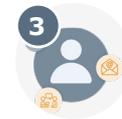

Putting the SUM Values Into Practice: Stakeholder Engagement Process and Project Summary Report

## Activities Section

This section contains instructions for group-based activities (each corresponding to a section in the Key Concepts). These activities are intended to increase understanding of Key Concepts by using them.

*Case studies within the AI Ethics and Governance in Practice workbook series are grounded in public sector use cases, but do not reference specific AI projects.*

**Relating to Values**

Demystify AI ethics by building a common vocabulary of the SUM Values grounded on your group's personal and collective relationship to each of them.

**Stakeholder Analysis**

Practise identifying vulnerable stakeholders by anticipating specific project's impacts on individuals and communities.

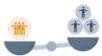 **Stakeholder Prioritisation**

Practise evaluating stakeholder prioritisation.

**Positionality Reflection**

Practise reflecting on your team's positionality with respect to case-specific stakeholders.

**Establishing an Engagement Objective**

Practise tailoring stakeholder engagement objectives to the needs of specific projects.

**Note for Facilitators**

Additionally, you will find facilitator instructions (and where appropriate, considerations) required for facilitating activities and delivering capacity-building workshops.

AI Sustainability in Practice Part One:
Foundations for Sustainable AI Projects

# Key Concepts

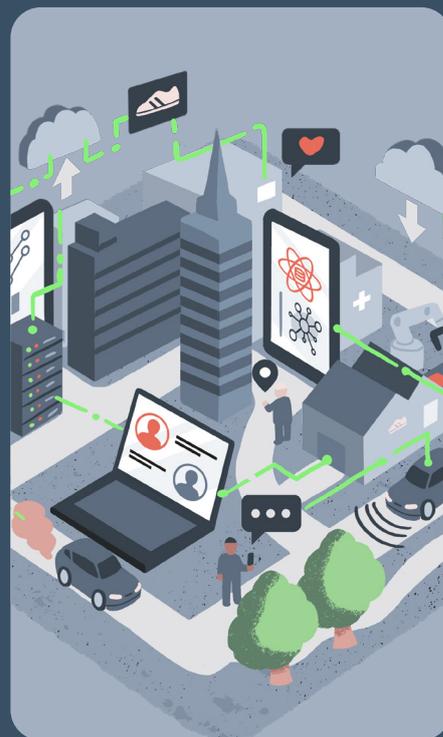

# Introduction to Sustainability: The SUM Values

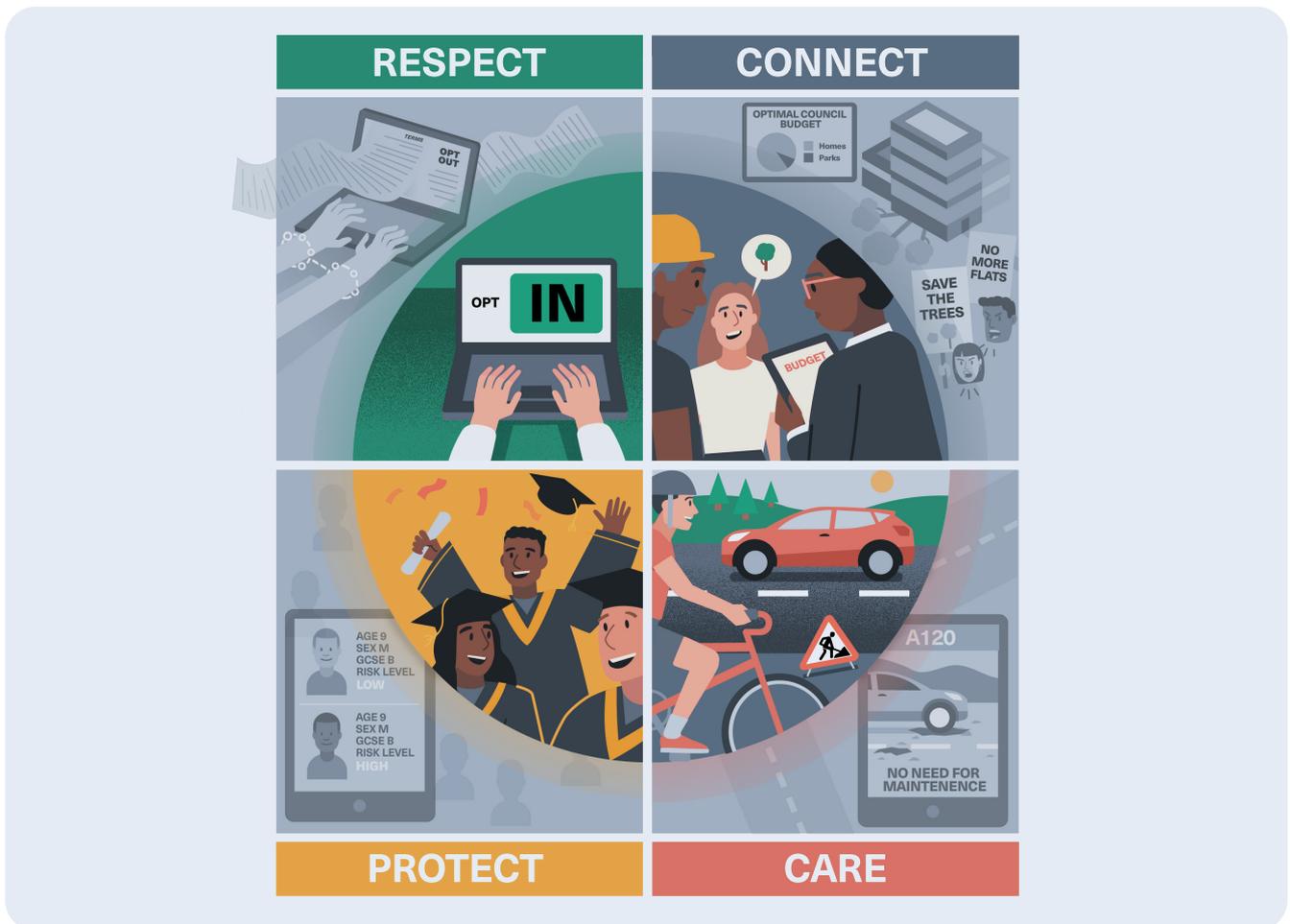

The challenge of creating a culture of responsible innovation begins with building an accessible moral vocabulary that will allow team members to navigate the ethical stakes of AI projects. In the field of AI ethics, this moral vocabulary draws primarily on two traditions of moral thinking: (1) bioethics and (2) human rights. Bioethics is the study of the ethical impacts of biomedicine and the applied life sciences. Human rights discourse draws its inspiration from the UN Declaration of Human Rights and other international instruments such as the European Convention on Human Rights, the European Social Charter, and the International Covenant on Civil and Political Rights. It is anchored in a set of universal principles that build upon the idea that all humans have an equal moral status as bearers of intrinsic human dignity.[4] We start this section by introducing bioethics and human rights as departure points of AI/ML ethics.

The human rights aspect of the design, development, and implementation of AI/ML systems is particularly important. In the UK, these fundamental rights and freedoms have been codified in the Human Rights Act (1998) and in the European Convention on Human Rights (1953).

Most generally, human rights are the basic rights and freedoms that are possessed by every person in the world from cradle to grave. They preserve and protect the inviolable dignity of each individual regardless of their:

race   ethnicity   gender   age

sexual orientation   class   religion

disability status   language   nationality

or any other ascribed characteristic

**KEY CONCEPT**

### Normativity

In the context of practical ethics, "normativity" means that a given concept, value, or belief puts a moral demand on one's practices. Such a concept, value, or belief indicates what one "should" or "ought to" do in circumstances where that concept, value, or belief applies. For example, if I hold the moral belief that helping people in need is a good thing, then, when confronted with a sick person in the street who requires help, I *should* help them. My belief puts a normative demand on me to act in accordance with what it is indicating that I *ought to* do, namely, to come to the person's aid.

These fundamental rights and freedoms create obligations that bind civil servants and governments to respecting, protecting, and fulfilling human rights. When these duties are not fulfilled, individuals are entitled to legal remedies and redressal of any human rights violations.[5]

Bioethics largely stresses the normative values that underlie the safeguarding of individuals in instances where technological practices affect their interests and wellbeing. The main principles of bioethics include:

- respecting the autonomy of the individual;

- protecting people from harm;

- looking after the wellbeing of others; and

- treating all individuals equitably and justly.

Human rights mainly focus on the set of social, political, and legal entitlements that are due to all human beings under a universal framework of juridical protection and the rule of law. The main tenets of human rights include:

- the entitlement to equal freedom and dignity under the law;

- the protection of civil, political, and social rights;

- the universal recognition of personhood; and

- the right to free and unencumbered participation in the life of the community.

# Origins of the SUM Values:
## Drawing Principles From Real-World Harms

Values are historically relative and contextually situated. This is a barrier that must be dealt with by guidance on ethical values intended to support responsible AI research and innovation.[6] It is reasonable to assume that, amid the moral and cultural plurality of modern social life, no fixed or universally accepted list of ethical values could definitively provide such a common starting point.[7]

As a result, researchers in AI/ML have had to take a more pragmatic and empirically driven position in proposing basic values. This proposal begins by considering the set of real-world problems posed by the use of AI/ML systems and data-driven technologies. These hazards include the potential loss of human agency and social connection in the wake of expanding automation, harmful outcomes that may result from the use of poor quality data or poorly designed, misused, or abused systems, and the possibility that entrenched societal dynamics of bias and discrimination will be perpetuated or even augmented by data-driven AI/ML technologies that tend to reinforce existing social and historical patterns.

**KEY CONCEPT**

### Bioethics and Human Rights in Context

The departure point of AI/ML ethics in the dynamics of real-world harm sheds some light on the appeal of bioethics and human rights. The principles that have emerged from both traditions found their origins in moral claims that have responded directly to tangible, technologically inflicted harms and atrocities. That is, both traditions emerged out of concerted public acts of resistance against violence done to disempowered or vulnerable people.[8]

Human rights has its origins in efforts to redress the barbarisms and genocides of the mid-twentieth century. The emergence of bioethics can be tracked to the public exposure in the 1960s and 1970s to several atrocities of human experimentation (such as the infamous Tuskegee syphilis experiment). In the latter instances, it was discovered that members of vulnerable or marginalised social groups had been subjected to the injurious effects of institutionally run biomedical experiments without having knowledge of or giving consent to their participation.[9] [10]

**KEY CONCEPT**

### Human Agency

Human agency is the capacity 'to act on one's own volition, to make one's own choices about how to live and flourish, and to freely pursue one's own life path'.[11]

# Real-World Hazards Posed by the Use of AI/ML Technologies

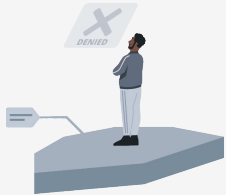

### 1. Loss of Autonomy

Integrating automated systems into social services may undermine human agency and autonomy, challenging individuals' self-determination and their ability to make decisions about their lives. Individuals may be disempowered or their sense of privacy may be violated. Algorithmic nudging techniques that treat people as malleable objects of prediction may promote the displacement of individual agency and the degradation of the conditions needed for the successful exercise of human judgment, moral reasoning, and practical rationality.

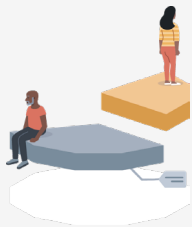

### 2. Loss of Interpersonal Connection and Empathy

Consequences of automation may also be potentially polarising and dehumanising. Opaque algorithmic models used by social media and digital platforms may create echo chambers and filter bubbles, undermining informational plurality and meaningful interpersonal dialogue. Individuals, who are subjected to automated decisions may feel that they have been 'reduced to statistics', and crucial human connection, association, and empathy may be lost.[12]

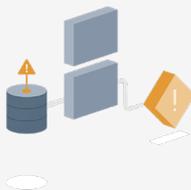

### 3. Poor Quality or Hazardous Outcomes

Algorithmic models are only as good as the data on which they are trained, tested, and validated (commnonly called 'garbage in, garbage out').[13] [14] [15] Inaccuracies, measurement errors, and sampling biases across data collection and recording can taint datasets. Using poor quality data could have grave consequences for individual wellbeing and public welfare. Likewise, poorly designed, misused, or abused AI systems may cause harmful outcomes. A powerful "frontier" AI model, for instance, could be misused to generate cybersecurity threats, bioweapons, or scaled disinformation.

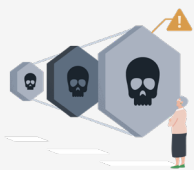

### 4. Bias, Inequity, and Discrimination

AI/ML models draw insights from existing patterns of data that can contain historical biases and inequities. Such models, in fact, make accurate out-of-sample predictions by replicating these patterns of the past—regardless of whether these patterns are inequitable or discriminatory.[16] [17] [18] This means such models could reproduce and amplify injustice.

Anchoring the foundations of AI ethics in real-world social harms provides a useful strategy for ethical deliberation. It has enabled the scope of the values and ethical concerns that underwrite responsible practices in the design, development and deployment of AI and data-driven systems to be informed by the actual risks posed by their use, as illustrated below:

| Risks that Emerge From the Use of AI/ML Technologies | Related Ethical Implications and Concerns |
|---|---|
| 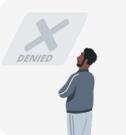 Loss of autonomy | 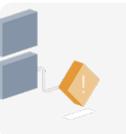 **Agency** Human agency, dignity, and individual flourishing |
| 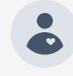 Loss of interpersonal connection and empathy | 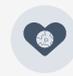 **Interaction** Solidarity, communication, and integrity of social interaction |
| 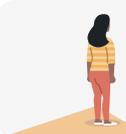 Poor quality or hazardous outcomes | 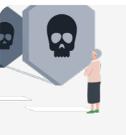 **Wellbeing** Individual, communal, and biospheric wellbeing |
| 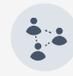 Bias, injustice, inequality, and discrimination | 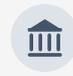 **Justice** Justice, equity, and the common good |

Over the past several years, dozens of AI ethics frameworks have emerged from various sectors such as academia, government, industry, and the third sector. These frameworks have converged around common values that aim to address these harms and hazards to society. The principles of bioethics and human rights have been instrumental in laying the groundwork for the development of such frameworks.

# The SUM Values in Focus
## Respect, Connect, Care, and Protect

The SUM Values follow this hazards-responsive approach by integrating concepts from bioethics and human rights. In particular, they do this by focusing on the most critical elements of these concepts to address the ethical and social issues that arise from the potential misuse, abuse, poor design, or unintended harmful consequences of AI systems. The SUM Values respond directly to the ethical implications and concerns related to these harms.

| Risks that Emerge From the Use of AI/ML Technologies | Related Ethical Implications and Concerns | SUM Values |
|---|---|---|
| Loss of autonomy | 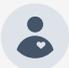 **Agency** | 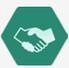 **Respect** the dignity of individual persons |
| Loss of interpersonal connection and empathy | 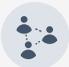 **Interaction** | 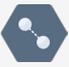 **Connect** with each other sincerely, openly, and inclusively |
| Poor quality or hazardous outcomes | 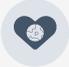 **Wellbeing** | 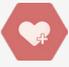 **Care** for the wellbeing of each and all |
| Bias, injustice, inequality, and discrimination | 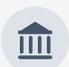 **Justice** | 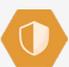 **Protect** the priorities of social values, justice, and the public interest |

The SUM Values are values that **support**, **underwrite**, and **motivate** a responsible innovation ecosystem. They provide an accessible framework of criteria for considering, assessing, and deliberating on the potential impacts and the **ethical permissibility** of a prospective AI project. They are meant to be utilised as **guiding values throughout the innovation lifecycle**: from the preliminary steps of project evaluation, planning, and problem formulation, through processes of design, development, and testing, to the stages of implementation and reassessment. Adopting common values from the outset enables reciprocally respectful, sincere, and open dialogue about ethical challenges by helping to create a shared vocabulary for informed dialogue and impact assessment. This common starting point also facilitates avenues for discussion and deliberation on how to balance values, especially when these values may come into conflict with one another.

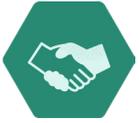

# Respect
## The Dignity of Individual Persons

- Ensure their ability to make free and informed decisions about their own lives.

- Safeguard their autonomy, their power to express themselves, and their right to be heard.

- Secure their capacities to make well-considered and independent contributions to the life of the community.

- Value the uniqueness of their aspirations, cultures, contexts, and forms of life.

- Secure their ability to lead a private life in which they are able to intentionally manage the transformative effects of the technologies that may influence and shape their development.

- Support their ability to flourish, to fully develop themselves, and to pursue their passions and talents according to their own freely determined life plans.

### Ethical Concerns

- Dignity, autonomy, agency, and authority of persons

- Self-realisation and flourishing of individuals

### Example

Psychiatrists use an AI system to support their assessments by analysing the patient's speech and offering recommendations for follow-up questions or actions. If a patient does not feel comfortable with the system analysing their speech automatically, there is the risk of undermining their autonomy and harming their dignity.

### Related Human Rights and Fundamental Freedoms

- The right to human dignity

- The right to life

- The right to liberty and security

- The right to respect for private and family life and the protection of personal data

- Freedom of thought, conscience, and religion

- Freedom of expression and opinion

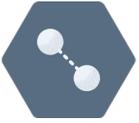

# Connect
## With Each Other Sincerely, Openly, and Inclusively

- Safeguard the integrity of interpersonal dialogue, meaningful human connection, and social cohesion.

- Prioritise and encourage diversity, participation, inclusion, and consideration of all voices across the AI/ML project lifecycle.

- Encourage all voices to be heard and all opinions to be weighed seriously and sincerely throughout the production and use lifecycle.

- Use AI innovations to enable bonds of interpersonal solidarity to form and individuals to be socialised and recognised by each other.

- Use AI technologies to foster the capacity to connect so as to reinforce trust, empathy, reciprocal responsibility, and mutual understanding.

### Ethical Concerns

- Integrity of interpersonal relationships

- Solidarity

- Participation-based innovation and stakeholder inclusion

### Related Human Rights and Fundamental Freedoms

- The right to diverse and reliable information and access to a plurality of ideas and perspectives

- The right to participate in the conduct of public affairs and good governance

- Freedom of assembly and association

### Example

If the use of the speech analysis AI system undermines the quality of the patient's interaction with the psychiatrist, there is the risk of harming the integrity of the patient-doctor relationship and undermining meaningful human connection.

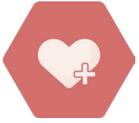

# Care
## For the Wellbeing of Each and All

- Design and deploy AI systems to foster and to cultivate the welfare of all stakeholders whose interests are affected by their use.

- Do no harm with these technologies and minimise the risks of their misuse or abuse.

- Prioritise the safety and the mental and physical integrity of people when scanning horizons of technological possibility and when conceiving of and deploying AI applications.

### Ethical Concerns

- Beneficence, safety, and non-harm

- Stewardship of individual, communal, and biospheric wellbeing

### Example

If the AI system that analyses patient's speech is trained on poor quality data or unrepresentative datasets, it may perform badly and make errors that harm the patient's wellbeing.

### Related Human Rights and Fundamental Freedoms

- The right to life

- The right to physical, mental, and moral integrity

- Environmental sustainability as the foundation for the enjoyment of all rights and freedoms

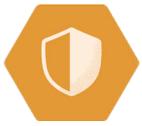

# Protect

## The Priorities of Social Values, Justice, and the Public Interest

- Treat all individuals equally and protect social equity.

- Use digital technologies as an essential support for the protection of fair and equal treatment under the law.

- Prioritise social welfare, public interest, and the consideration of the social and ethical impacts of innovation in determining the legitimacy and desirability of AI technologies.

- Use AI to empower and to advance the interests and wellbeing of as many individuals as possible.

- Reflect on how the design, development, and use of AI systems may affect the interests of others around the globe, future generations, and the biosphere as a whole.

### Ethical Concerns

- Justice and equity

- Prioritisation of the public interest and common good

### Example

If the AI system used to support assessment and diagnosis of a psychiatric patient is trained with data from a specific and unrepresentative subset of the population and then used on the wider population, it could discriminate against certain subgroups and cause unjust outcomes.

### Related Human Rights and Fundamental Freedoms

- The right to non-discrimination

- The right to equality before the law

- The right to an effective remedy for violation of rights and freedoms

- The right to a fair trial and due process

- The right to judicial independence and impartiality

- The right to equality of arms and opportunity in a court of law

# Putting the SUM Values Into Practice

**Stakeholder Engagement Process and Project Summary Report**

The SUM Values provide your project team with a shared vocabulary to assess the ethical permissibility of your project and its potential impacts. To engage in this kind of anticipatory reflection, your project team first needs to gain a contextually informed understanding of the social environment and human factors that may be impacted by, or may impact, the tool or model you are planning to develop. However, there are often gaps between project teams and these contexts. The Stakeholder Engagement Process (SEP) provides the tools to facilitate proportionate engagement and input from stakeholders with more understanding of these contexts. This process should initially be conducted during the Project Planning step within the Design Phase of the AI lifecycle. It should then be revisited at each iteration of the Stakeholder Impact Assessment (SIA).

Part one of this workbook will take a deep dive into the Stakeholder Engagement Process, that is, the process by which the project team establishes engagement objectives and designs engagement methods. AI Sustainability in Practice Part Two will cover Stakeholder Impact Assessments.

---

**KEY CONCEPT**

### Stakeholder

Scholars and practitioners from areas as diverse as public policy, land use, environmental and natural resource management, international development, and public health have offered many different definitions of 'stakeholders' over the past several decades.[19] Even so, these definitions have converged around a few common characteristics. Stakeholders are individuals or groups that:

1. have interests or rights that may be affected by the past, present, and future decisions and activities of an organisation;

2. may have the power or authority to influence the outcome of such decisions and activities; and/or

3. have relevant characteristics that put them in positions of advantage or vulnerability with regard to those decisions and activities.

The Stakeholder Engagement Process focuses on impacted communities and groups.

---

**KEY CONCEPT**

### Stakeholder Impact Assessment

A tool that creates a procedure for, and a means of, documenting the collaborative evaluation and reflective anticipation of the possible harms and benefits of AI innovation projects.

# Three Steps of Stakeholder Engagement Process (SEP)

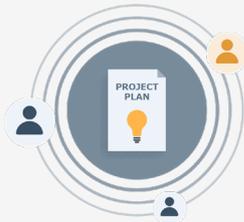

**1.  Preliminary Project Scoping and Stakeholder Analysis**

Outline key project components. Identify individuals or groups who may be affected by, or may affect, your innovation project. Scope potential stakeholder impacts. Evaluate the salience and contextual characteristics of identified stakeholders.[20]

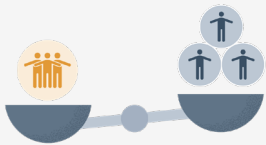

**2.  Positionality Reflection**

Evaluate team positionality as related to that of stakeholders. Consider strengths and limitations presented by team positionality.[21] [22] [23]

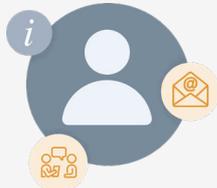

**3.  Stakeholder Engagement Objectives and Methods**

Establish engagement objectives that enable the appropriate degree of stakeholder engagement and co-production in project evaluation, and methods that support the achievement of defined objectives.[24] [25]

Each of these activities should be documented and utilised to create a Project Summary (PS) Report. A Project Summary Report is comprised of three components reflecting the SEP:

1. a preliminary project scoping and stakeholder analysis;

2. a positionality reflection; and

3. an overview of established stakeholder objectives and methods.

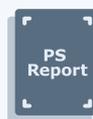

The **Project Summary Report template** can be found on page 39.

The SEP is not a one-off activity, but rather should be seen as an ongoing process that occurs throughout the project lifecycle.[26] The initial SEP takes place internally (i.e. within the organisation or team) during the design phase of the project workflow and is informed by available organisational scoping and planning documents and by desk-based research. These resources are used to develop an initial PS Report that is revisited during each subsequent SIA. The SIA facilitates the iterative evaluation of the social impact and sustainability of individual AI projects, as well as the corroboration of these impacts in dialogue with stakeholders, when appropriate. After each SIA, the PS Report is updated to reflect up-to-date stakeholder analyses, positionality reflections, and engagement objectives and methods established for the following SIA. Each iteration includes re-evaluating which stakeholders to engage.

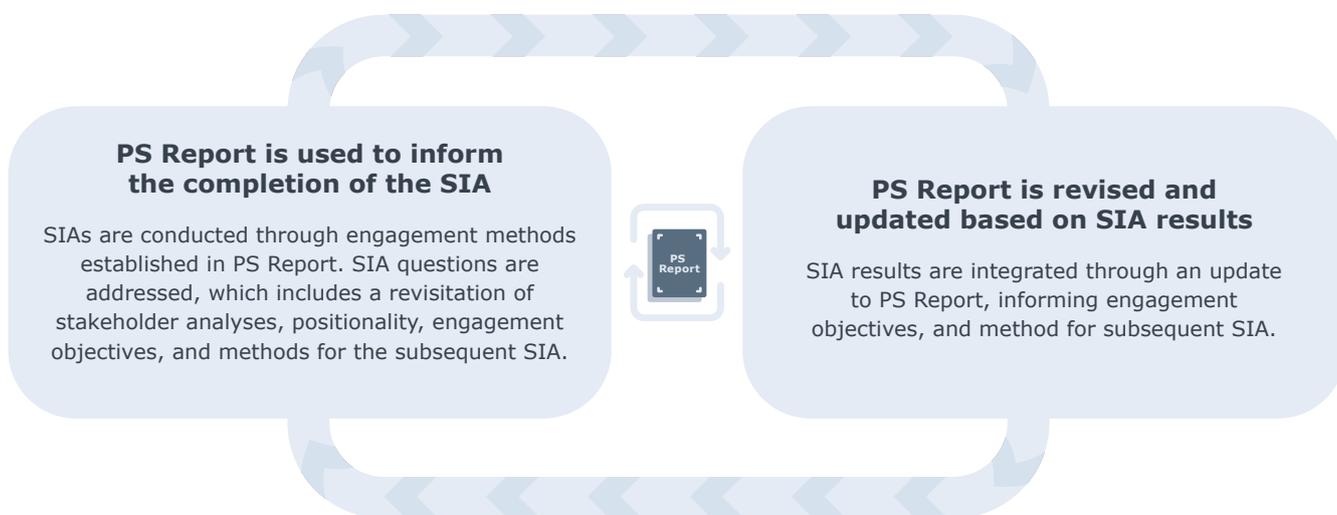

**PS Report is used to inform the completion of the SIA**

SIAs are conducted through engagement methods established in PS Report. SIA questions are addressed, which includes a revisitation of stakeholder analyses, positionality, engagement objectives, and methods for the subsequent SIA.

**PS Report is revised and updated based on SIA results**

SIA results are integrated through an update to PS Report, informing engagement objectives, and method for subsequent SIA.

It is important to revise and update your PS Report to ensure that engagement objectives and methods for conducting SIAs continue to reflect the perspectives and interests of salient stakeholders across the AI lifecycle. This relationship will be discussed in depth in AI Sustainability in Practice Part Two.

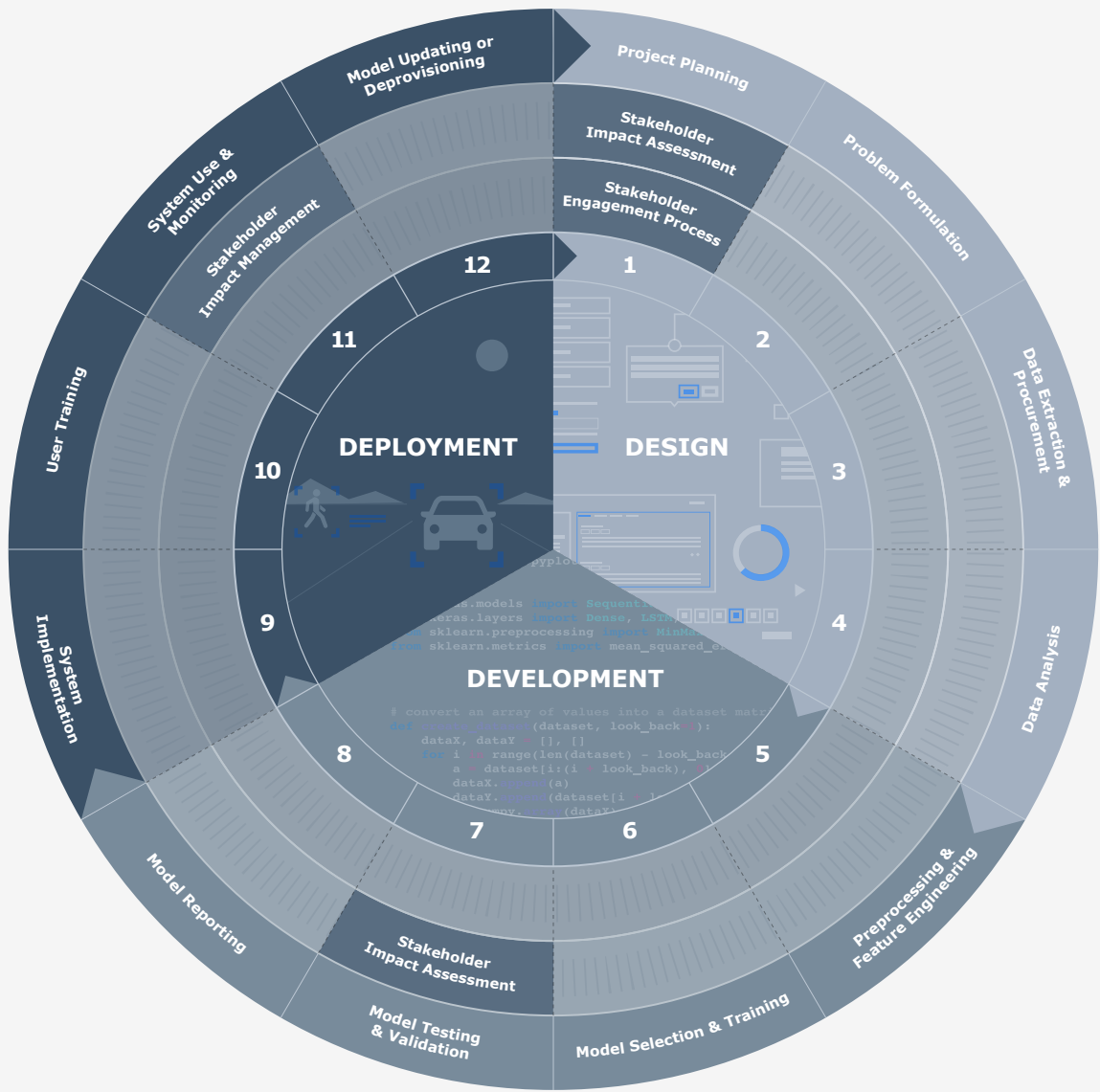

Project Planning
Stakeholder Impact Assessment
Stakeholder Engagement Process
Problem Formulation
Data Extraction & Procurement
Data Analysis
Preprocessing & Feature Engineering
Model Selection & Training
Model Testing & Validation
Stakeholder Impact Assessment
Model Reporting
System Implementation
User Training
System Use & Monitoring
Model Updating or Deprovisioning
Stakeholder Impact Management

DESIGN
DEVELOPMENT
DEPLOYMENT

1
2
3
4
5
6
7
8
9
10
11
12

# Preliminary Project Scoping and Stakeholder Analysis

Preliminary Project Scoping and Stakeholder Analysis is the first activity within the SEP process. It involves four sub-steps (details of each of these are included in the Project Summary Report template on ):

**Sub-step 1**

## Outlining Project, Use Context, Domain, and Data

Outline a high-level description of the prospective system, the contexts in which it will be used, the domain in which it will operate, and the data on which it will be trained. During this initial project scoping activity, you should draw on organisational documents (i.e. the project business case, proof of concept, or project charter), project team collaboration, and desk research (if necessary) to complete the description.

**Sub-step 2**

## Identifying Stakeholders

Building on this contextual understanding, identify who may be affected by, or may affect, your AI project. While doing this, consider protected characteristics and vulnerability factors.

---

**KEY CONCEPT**

**Protected Characteristics**
These are legally protected identity traits that could increase a person's vulnerability to abuse, adverse impact, or discrimination. Under the Equality Act 2010, nine characteristics are protected against discrimination. These characteristics are: age, disability, gender reassignment, marriage and civil partnership, pregnancy and maternity, race, religion or belief, sex, sexual orientation. Stakeholders with protected characteristics may require additional protection or assistance with respect to project impacts.

**Vulnerability Factors**
These are elements beyond the nine legally protected characteristics which could increase a persons' vulnerability to project impacts. These factors may include distinct facets of a singular person's identity, circumstances, or contexts which may expose them to being jeopardised by a project.

---

**Sub-step 3**

**Scoping Potential Stakeholder Impacts**

Carry out a preliminary evaluation of the potential impacts of the prospective AI system on affected individuals and communities. At this initial stage of reflection, members of your project team should review the SUM Values, and the corresponding ethical concerns, rights and freedoms, and then consider which of these could be impacted by the design, development, and deployment of the prospective AI system and how.

**Sub-step 4**

**Analysing Stakeholder Salience**

Assess the relevance of each identified stakeholder group to your project and to its use contexts. Assess the relative interests, rights, vulnerabilities, and advantages of identified stakeholders as these interests, rights, vulnerabilities, and advantages may be impacted by, or may impact, the AI system your team is planning to develop and deploy. When identifying stakeholders, your team should also consider organisational stakeholders, whose input will likewise strengthen the openness, inclusivity, and diversity of your project.

---

**KEY CONCEPT**

**Salience**
The degree to which an impacted stakeholder:

1. has legitimate interests, rights, or vulnerabilities that could be affected by the project;

2. can exercise influence on the project; or

3. has claims on the project and its outcomes that demand urgent attention.[27]

# Determining a Proportionate Approach to Stakeholder Involvement

Stakeholder analyses may be carried out in a variety of ways that involve more or less stakeholder involvement. This spectrum of options ranges from analyses carried out exclusively by a project team without active community engagement to analyses built around the inclusion of community-led participation and co-design from the earliest stages of stakeholder identification. The degree of stakeholder involvement will vary from project to project based upon a preliminary assessment of the potential risks and hazards of the model or tool under consideration.

Low-stakes AI applications that are not safety-critical, do not directly impact the lives of people, and do not process potentially sensitive social and demographic data may need less proactive stakeholder engagement than high-stakes projects. You and your project team will need to carry out an initial evaluation of the scope of the possible risks that could arise from your project and of the potential hazards it poses to affected individuals and groups. You will have to apply reasonable assessments of the dangers posed to individual wellbeing and public welfare in order to formulate proportionate approaches to stakeholder involvement.

Regardless of the potential impacts of a project, involving affected individuals and communities in stakeholder analysis (and, later, in stakeholder impact assessment) should, in all cases, be a significant consideration. Stakeholder involvement ensures that your project will possess an appropriate degree of public accountability, transparency, legitimacy, and democratic governance, and it recognises the important role played in this by the inclusion of the voices of all affected individuals and communities in decision-making and policy articulation processes.[28] [29] [30]

In addition to providing these important supports for building public trust, stakeholder involvement can help to strengthen the objectivity, reflexivity, reasonableness, and robustness of the choices your project team makes across the project lifecycle.[31] This is because the inclusion of a wider range of perspectives (especially of those who are most marginalised) can enlarge a project team's purview and expand its domain knowledge as well as its understanding of the public's needs.[32] [33] It can likewise unearth potential biases that may arises from limiting the standpoints that inform decision-making to those of team members.

Public engagement and community involvement, however, are only one part of the measures your team needs to take to ensure the objectivity, reflexivity, reasonableness, and robustness of its stakeholder analysis, impact assessment, and decision-making more generally. Apart from outward-facing community participation, processes of inward-facing reflection should also inform the way your team approaches to these challenges.

# Engagement and Reflection Cycle of Sustainable AI Innovation

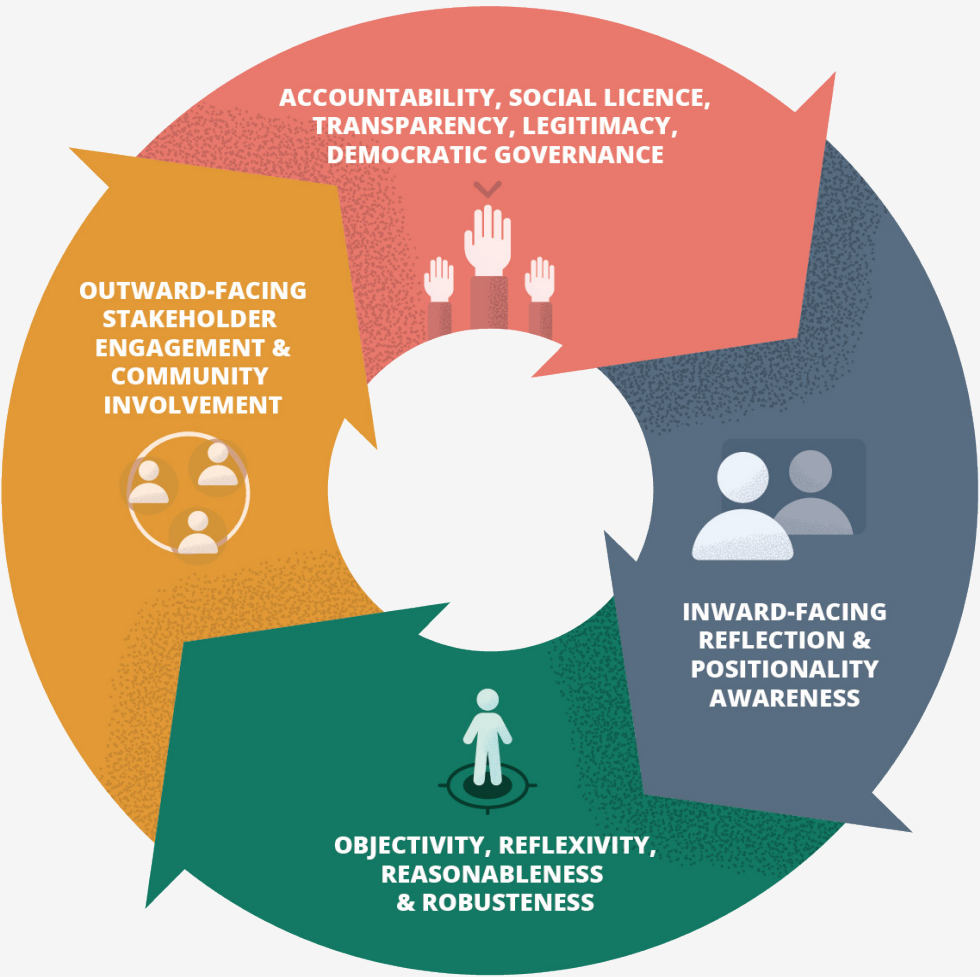

ACCOUNTABILITY, SOCIAL LICENCE, TRANSPARENCY, LEGITIMACY, DEMOCRATIC GOVERNANCE

OUTWARD-FACING STAKEHOLDER ENGAGEMENT & COMMUNITY INVOLVEMENT

INWARD-FACING REFLECTION & POSITIONALITY AWARENESS

OBJECTIVITY, REFLEXIVITY, REASONABLENESS & ROBUSTNESS

# Positionality Reflection

All individual human beings come from unique places, experiences, and life contexts that have shaped their thinking and perspectives. Reflecting on these can help us understand how our viewpoints might differ from those around us, especially those who have diverging cultural and socioeconomic backgrounds and life experiences.[34] Identifying and probing these differences can enable individuals to better understand how their own backgrounds, for better or worse frame:

- the way they see others;
- the way they approach and solve problems; and
- the way they carry out research and engage in innovation.

By undertaking such efforts to recognise social position and differential privilege, individuals may gain a greater awareness of their own personal biases and unconscious assumptions. This then can enable them to better discern the origins of these biases and assumption and to confront and challenge them in turn.[35]

Social scientists have long referred to this kind of self-locating reflection as "positionality".[36] [37] [38] When team members take their own positionalities into account, and make them explicit, they can better grasp how the influence of their respective social and cultural positions creates strengths and limitations. On the one hand, one's positionality—with respect to characteristics like ethnicity, race, age, gender, socioeconomic status, education and training levels, values, geographical background, etc.—can have a positive effect on an individual's contributions to an innovation project. The uniqueness of each person's lived experience and standpoint can play a constructive role in introducing insights and understandings that other team members do not have.[39] On the other hand, one's positionality can assume a harmful role when hidden biases and prejudices that derive from a person's background, and from power imbalances and differential privileges, creep into decision-making processes undetected.[40]

**When taking positionality into account, your team members should reflect on their own positionality matrix. They should ask:**

- To what extent do my personal characteristics, group identifications, socioeconomic status, educational, training, and work background, team composition, and institutional frame represent sources of power and advantage or sources of marginalisation and disadvantage?

- How does this positionality influence my (and my team's) ability to identify and understand affected stakeholders and the potential impacts of my project?

**Several other questions must be asked to respond to these two:**

- How do I identify?
- How have I been educated and trained?
- What does my institutional context and team composition look like?
- What is my socioeconomic history?

### How do I Identify?

Age, race and ethnicity, disability status, religion, gender, sexuality, marital status, parental status, linguistic background.

**Personal Characteristics and Group Identifications[41] [42]**

### How Have I Been Educated and Trained?

Schools attended, level of education, opportunities for advancement and professional development, employment history.

**Education, Training, and Work Background[43] [44] [45] [46]**

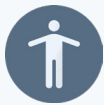

# Positionality Matrix

To what extent do my personal characteristics, group identifications, socioeconomic status, educational, training, and work background, team composition, and institutional frame represent sources of power and advantage or sources of marginalisation and disadvantage? How does this positionality influence my (and my team's) ability to identify and understand affected stakeholders and the potential impacts of my project?

**Institutional Frame and Team Composition[47] [48]**

**Socioeconomic Status[49]**

### What Does My Institutional Context and Team Composition Look Like?

Authority structure within my project team, wider policy ownership and power hierachies in my organisation, levels of decision-making autonomy, opportunities to voice concerns and objections, team diversity, culture of inclusion or exclusion.

### What Is My Socioecomionic History?

Socioeconomic status growing up, social mobility over time, present status, socioeconomic aspirations.

---

**The following resources can aid in further understanding marginalisation and positionality:**

Understanding Oppression: Strategies in Power and Privilege by Leticia Nieto and Margot F. Boyer

Understanding Oppression Part 2: Skill Sets for Targets by Leticia Nieto and Margot F. Boyer

Understanding Oppression Part 3: Skill Sets for Agents by Leticia Nieto and Margot F. Boyer

Diversity, Equity, and Inclusion Practice Guide by 501 Commons

See also Stakeholder Analysis, Positionality, and Engagement resources and literature contained within the Bibliography and Further Readings section on  of this workbook.

# Determining Stakeholder Engagement Objectives for Stakeholder Impact Assessments

Stakeholder engagement may be carried out in a variety of ways that involve more or less stakeholder involvement. This spectrum of options ranges from analyses carried out exclusively by a project team without active community engagement to analyses built around the inclusion of community-led participation and co-design from the earliest stages of stakeholder identification.

The objectives of engagement will vary from project to project and will depend on the following factors:

| Factors Determining the Objectives and Methods of Engagement | |
|---|---|
| 1   **Team-Based Assessments of Risks of Adverse Impacts** | • Assessment of how to make stakeholder involvement proportionate to the scope of a project's potential risks and hazards. |
| 2   **Team-Based Assessments of Positionality** | • Evaluation of team positionality—for instance, cases where the identity characteristics of team members do not sufficiently reflect or represent significantly impacted groups. How can the project team "fill the gaps" in knowledge, domain expertise, and lived experience through stakeholder participation? |

When weighing these two factors, your team should prioritise the establishment of a clear and explicit stakeholder engagement objective. The stakeholder engagement objective will enable appropriate degree of stakeholder agency in project evaluation and oversight processes, that is, the control stakeholders have over the actions in project evaluation and oversight processes and their consequences. Documenting the stakeholder engagement objective is crucial because all stakeholder engagement processes can run the risk either of being cosmetic tools employed to legitimate projects without substantial and meaningful participation or of being insufficiently participative, i.e. of being one-way information flows or nudging exercises that serve as public relations instruments.[50] To avoid such hazards of superficiality, your team should shore up its proportionate approach to stakeholder engagement with deliberate and precise goal-setting.

Your stakeholder engagement objective will entail choosing from a spectrum of engagement options with varying degree of stakeholder agency (informing, partnering, consulting, empowering) that equip your project with a level of engagement which meets team-based assessments of risks of adverse impacts and positionality.

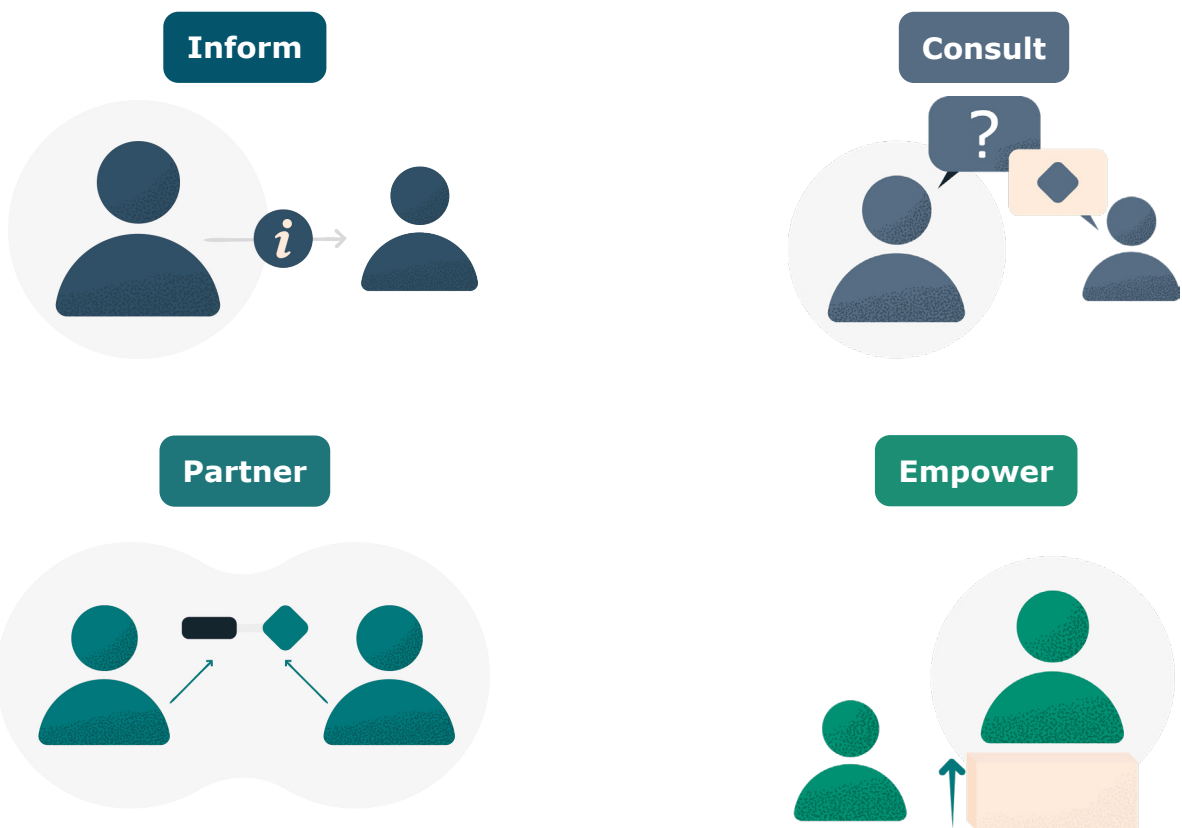

| Stakeholder Engagement Objective | Level of Agency |
|---|---|
| **Inform** | |
| Stakeholders are made aware of decisions and developments. | LOW<br>Stakeholders are considered information subjects rather than active agents. |
| **Consult** | |
| Stakeholders can voice their views on pre-determined areas of focus, which are considered in decision-making. | LOW<br>Stakeholders are included as sources of information input under narrow, highly controlled conditions of participation. |
| **Partner** | |
| Stakeholders and teams share agency over the determination of areas of focus and decision-making. | MODERATE<br>Stakeholders exercise a moderate level of agency in helping to set agendas through collaborative decision-making. |
| **Empower** | |
| Stakeholders are engaged with as decision-makers and are expected to gather pertinent information and be proactive in co-operation. | HIGH<br>Stakeholders exercise a high level of agency and control over agenda-setting and decision-making. |

*Table informed by The Local Government Association's Councillor's workbook on neighbourhood and community engagement.*

## Example

The local authority in a city with growing commercial and university sectors is looking to develop an AI system that can optimise public transportation routes and schedules, reduce congestion, and improve overall mobility and transport for residents. To inform the design of the AI system, the local authority has decided to engage with different stakeholder groups. This will help them understand their needs and expectations, as well as identify any potential impact of the proposed tool. The local authority anticipates that, if sufficient attention is not put on mitigating risks from the AI system, it could have negative impacts on the local community. For this reason, they consider that the engagement objectives of Partner and Empower could be best suited for a medium to high risk project.

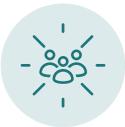

**Partner**

The local authority conducts **focus groups** with residents representative of the community and diverse in terms of age, gender, ethnicity, occupation, and socioeconomic status. They will share their views and experiences of mobility in the city, and help determine the key areas of focus for potential impacts of the AI system.

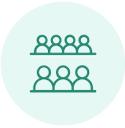

**Empower**

The local authority will recruit a group of people who are representative of the local community to be part of a **citizens' jury**. The jurors will gather evidence on the current mobility challenges and the potential impacts of the proposed AI system. They will deliberate and draft a roadmap for the design and development of the AI system, which includes a list of key areas of focus.

Both engagement options (focus groups and citizens' jury) serve as components of broader impact assessment processes.

# Determining Stakeholder Engagement Methods for Stakeholder Impact Assessments

Once you have established your engagement objective, you are in a better position to assess which method or methods of engaging stakeholders are most appropriate in conducting your Stakeholder Impact Assessments.

Determining appropriate engagement methods for conducting this process necessitates that you:

1. evaluate and accommodate stakeholder needs; and

2. pay attention to practical considerations of resources, capacities, timeframes, and logistics that could enable or constrain the realisation of your objective:[51] [52]

| Factors Determining the Objectives of Engagement | |
| --- | --- |
| **Evaluation and Accommodation of Stakeholder Needs** | • Identification of potential barriers to engagement. For instance, constraints on the capacity of vulnerable stakeholder groups to participate, difficulties in reaching marginalised, isolated, or socially excluded groups, and challenges to participation that are presented by digital divides or information and communication gaps between public sector organisations and impacted communities.[53] [54] [55]<br><br>• Identification of strategies to accommodate stakeholder needs, such as catering the location or media of engagement to difficult to reach groups. Provision of childcare, compensation, or transport to secure equitable participation.[56]<br><br>• Tailoring the information and educational materials to the needs of participants.[57]<br><br>• Consideration of engagement objectives. |
| **Practical Considerations of Resources, Capacities, Timeframes, and Logistics** | • The resources available for facilitating engagement activities.<br><br>• The timeframes set for project completion.<br><br>• The capacities of your organisation and team to properly facilitate public engagement.<br><br>• The stages of project design, development, and implementation at which stakeholders will be engaged. |

You and your team may face pitfalls when confronting any of these factors. For example, limits on available resources and tight timelines could be at cross-purposes with the degree of stakeholder agency that is recommended by team-based assessments of potential hazards and positionality limitations. Likewise, the chosen degree of appropriate citizen participation may be unrealistic or out-of-reach given the engagement barriers that arise from high levels of stakeholder needs. In these instances, you and your project team should take a deliberate and reflective approach to deciding on how to balance engagement objectives and stakeholder needs with practical considerations. And, you should make explicit the rationale behind your choices and document this.

<div style="background:#fce9e2;">

**This Range of Stakeholder Participation Options Lines up with the Following Engagement Methods:**[58]

</div>

| Engagement Method | Practical Strengths | Practical Weaknesses |
|---|---|---|
| 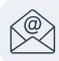 **Newsletters (Email)**<br><br>Regular emails (e.g. fortnightly or monthly) that contain updates, relevant news, and calls to action in an inviting format.<br><br>**Engagement Method:**<br>**Inform** | Can reach many people; can contain a large amount of relevant information; can be made accessible and visually engaging. | Might not reach certain portions of the population; can be demanding to design and produce with some periodicity; easily forwarded to spam/junk folders without project team knowing (leading to overinflated readership statistics). |
| 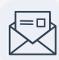 **Letters (Post)**<br><br>Regular letters (e.g. monthly) that contain the latest updates, relevant news and calls to action.<br><br>**Engagement Method:**<br>**Inform** | Can reach parts of the population with no internet or digital access; can contain large amount of relevant information; can be made accessible and visually engaging. | Might not engage certain portions of the population; slow delivery and interaction times hampers the effective flow of information and the organisation of further engagement. |

| Engagement Method | Practical Strengths | Practical Weaknesses |
|---|---|---|
| 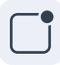 **App Notifications**<br><br>Projects can rely on the design of apps that are pitched to stakeholders who are notified on their phone with relevant updates.<br><br>**Engagement Method:**<br>**Inform** | Easy and cost-effective to distribute information to large numbers of people; rapid information flows bolster the provision of relevant and timely news and updates. | More significant initial investment in developing an app; will not be available to people without smartphones. |
| 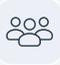 **Community Fora**<br><br>Events in which panels of experts share their knowledge on issues and then stakeholders can ask questions.<br><br>**Engagement Method:**<br>**Inform** | Can inform people with more relevant information by providing them with the opportunity to ask questions; brings community together in a shared space of public communication. | More time-consuming and resource-intensive to organise; might attract smaller numbers of people and self-selecting groups rather than representative subsets of the population; effectiveness is constrained by forum capacity. |
| 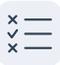 **Online Surveys**<br><br>Survey sent via email, embedded in a website, shared via social media.<br><br>**Engagement Method:**<br>**Consult** | Cost-effective; simple mass-distribution. | Risk of pre-emptive evaluative framework when designing questions; does not reach those without internet connection or computer/smartphone access. |

| Engagement Method | Practical Strengths | Practical Weaknesses |
|---|---|---|
| 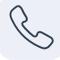 **Phone Interviews**<br><br>Structured or semi-structured interviews held over the phone.<br><br>**Engagement Method:**<br>Consult  Partner | Opportunity for stakeholders to voice concerns more openly. | Risk of pre-emptive evaluative framework when designing questions; might exclude portions of the populations without phone access or with habits of infrequent phone use. |
| 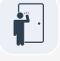 **Door-to-door Interviews**<br><br>Structured or semi-structured interviews held in-person at people's houses.<br><br>**Engagement Method:**<br>Consult  Partner | Opportunity for stakeholders to voice concerns more openly; can allow participants the opportunity to form connections through empathy and face-to-face communication. | Potential for limited interest to engage with interviewers; time-consuming; can be seen by interviewees as intrusive or burdensome. |
| 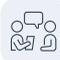 **In-person Interviews**<br><br>Short interviews conducted in-person in public spaces.<br><br>**Engagement Method:**<br>Consult  Partner | Can reach many people and a representative subset of the population if stakeholders are appropriately defined and sortition is used. | Less targeted; pertinent stakeholders must be identified by area; little time/interest to engage with interviewer; can be viewed by interviewees as time-consuming and burdensome. |
| 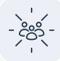 **Focus Groups**<br><br>A group of stakeholders brought together and asked their opinions on a particular issue. Can be more or less formally structured.<br><br>**Engagement Method:**<br>Consult  Partner | Can gather in-depth information; can lead to new insights and directions that were not anticipated by the project team. | Subject to hazards of groupthink or peer pressure; complex to facilitate; can be steered by dynamics of differential power among participants. |

| Engagement Method | Practical Strengths | Practical Weaknesses |
|---|---|---|
| 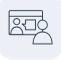 **Online Workshops**<br><br>Workshops using digital tools such as collaborative platforms.<br><br>**Engagement Method:**<br>`Consult` | Opportunity to reach stakeholders across regions; increased accessibility depending on digital access. | Potential barriers to accessing tools required for participation; potential for disengagement. |
| 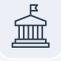 **Citizen Panel or Assembly**<br><br>Large groups of people (dozens or even thousands) who are representative of a town/region.<br><br>**Engagement Method:**<br>`Inform` `Partner`<br>`Empower` | Provides an opportunity for co-production of outputs; can produce insights and directions that were not anticipated by the project team; can provide an information base for conducting further outreach (surveys, interviews, focus groups, etc.); can be broadly representative; can bolster a community's sense of democratic agency and solidarity. | Participant roles must be continuously updated to ensure panels or assemblies remain representative of the population throughout their lifespan; resource-intensive for establishment and maintenance; subject to hazards of groupthink or peer pressure; complex to facilitate; can be steered by dynamics of differential power among participants. |
| 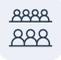 **Citizen Jury**<br><br>A small group of people (between 12 and 24), representative of the demographics of a given area, come together to deliberate on an issue (generally one clearly framed set of questions), over the period of 2 to 7 days.<br><br>**Engagement Method:**<br>`Inform` `Partner`<br>`Empower` | Can gather in-depth information; can produce insights and directions that were not anticipated by the project team; can bolster participants' sense of democratic agency and solidarity. | Subject to hazards of groupthink; complex to facilitate; risk of pre-emptive evaluative framework; small sample of citizens involved risks low representativeness of wider range of public opinions and beliefs. |

# Project Summary Report Template

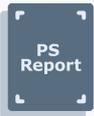 **Project Summary Report Template for:** Project Name

## Component 1: Project Scoping and Stakeholder Analysis

### 1. Outlining Project, Use Context, Domain, and Data

**a.** What AI system is being built and what type of product or service will it offer?

**b.** What benefits will the system bring to its users and customers, and will these benefits be widely accessible?

**c.** Which organisation(s)—yours, other suppliers, or other providers—are responsible for building this AI system?

**d.** Which parts or elements of the AI system, if any, will be procured from third-party vendors, suppliers, sub-contractors, or external developers?

**e.** Which algorithms, techniques, and model types will be used in the AI system? Provide links to technical papers where appropriate.

**f.** In a scenario where your project optimally scales, how many people will it impact, for how long, and in what geographic range (local, national, global)? Describe your rationale.

## 2. Use Context

**a.** What is the purpose of this AI system and in which contexts will it be used? Briefly describe a use-case that illustrates primary intended use.

........................................................

**b.** Is the AI system's processing output to be used in a fully automated way or will there be some degree of human control, oversight, or input before use? Describe.

........................................................

**c.** Will the AI system evolve or learn continuously in its use context or will it be static?

........................................................

**d.** To what degree will the use of the AI system be time-critical, or will users be able to evaluate outputs comfortably over time?

........................................................

**e.** What sort of out-of-scope uses could users attempt to apply the AI system, and what dangers may arise from this?

........................................................

## 3. Domain

**a.** In what domain(s) will this AI system operate?

........................................................

**b.** Which, if any, domain experts have been or will be consulted in designing, developing, and developing the AI system?

........................................................

## 4. Data

**a.** What datasets will be used to build this AI system?

..................................................

**b.** Will any data being used in the production of the AI system be acquired from a vendor or supplier? Describe.

..................................................

**c.** Will the data being used in the production of the AI system be collected for that purpose, or will it be re-purposed from existing datasets? Describe.

..................................................

## 5. Identifying Stakeholders

**a.** Who are the stakeholders (both individuals and social groups) that may be impacted by, or may impact, the project?

..................................................

**b.** Do any of these stakeholders possess sensitive or protected characteristics that could increase their vulnerability to abuse, adverse impact, or discrimination, or for reason of which they may require additional protection or assistance with respect to the impacts of the project? If so, what characteristics?

..................................................

**c.** Could the outcomes of this project present significant concerns to specific groups of stakeholders given vulnerabilities caused or precipitated by their distinct circumstances?

..................................................

**d.** If so, what vulnerability characteristics expose them to being jeopardised by project outcomes?

..................................................

## 6. Scoping Potential Stakeholder Impacts

**a.** How could each of the SUM Values and their associated ethical concerns values be impacted by the AI system we are planning to build?

### Respect
the dignity of individual persons

**Ethical Concerns:**

- Dignity, autonomy, agency, and authority of persons.

- Self-realisation and flourishing of individuals.

### Protect
the priorities of social values, justice, and the public interest

**Ethical Concerns:**

- Justice and equity.
- Prioritisation of the public interest and common good.

### Connect
with each other sincerely, openly, and inclusively

**Ethical Concerns:**

- Integrity of interpersonal connections.
- Solidarity.
- Participation-based innovation and stakeholder inclusion.

### Care
for the wellbeing of each and all

**Ethical Concerns:**

- Beneficence, safety, and non-harm.

**b.** If things go wrong in the implementation of our AI system or if it is used out-of-the-scope of its intended purpose and function, what harms could be done to stakeholders in relation to each of the SUM Values and their associated ethical concerns?

### Respect
the dignity of individual persons

**Ethical Concerns:**

- Dignity, autonomy, agency, and authority of persons.
- Self-realisation and flourishing of individuals.

.......................................................

### Protect
the priorities of social values, justice, and the public interest

**Ethical Concerns:**

- Justice and equity.
- Prioritisation of the public interest and common good.

.......................................................

### Connect
with each other sincerely, openly, and inclusively

**Ethical Concerns:**

- Integrity of interpersonal connections.
- Solidarity.
- Participation-based innovation and stakeholder inclusion.

.......................................................

### Care
for the wellbeing of each and all

**Ethical Concerns:**

- Beneficence, safety, and non-harm.

.......................................................

### 7. Analysing Stakeholder Salience

**a.** Which affected stakeholder groups are most likely to be positively impacted by the deployment of the system or tool? Which affected stakeholder groups are most likely to be negatively impacted?

......................................................................

**b.** Which affected stakeholder groups have the greatest needs in relation to potential benefits of the system or tool?

......................................................................

**c.** How might different affected stakeholder groups be differentially impacted by the system?

......................................................................

**d.** Are there any relevant power relations between these differentially impacted stakeholder groups that could affect the distribution of the prospective system's benefits and risks? Consider their relative advantages and disadvantages, and which affected stakeholders may have direct or indirect influence over the project and its outcomes.

......................................................................

**e.** Which affected stakeholder groups have existing influence within relevant communities, political processes, or in relation to the domain in which the system will be deployed? How could these dynamics of influence impact the distribution of the prospective system's benefits and risks?

......................................................................

**f.** Which affected stakeholder groups' influences are limited? How could these limitations impact the distribution of the prospective system's benefits and risks?

......................................................................

## Component 2: Positionality Reflection

**a.** How does the positionality of team members relate to that of affected stakeholders?

...................................................

**b.** How could your positionality as a team influence your evaluation of the potential negative and positive impacts of this project?

...................................................

**c.** In which ways could your positionality as a team limit your perspective when evaluating the impact of this project?

...................................................

**d.** How could your positionality as a team strengthen your perspective when evaluating the impact of this project? Consider overlapping identities and experience.

...................................................

**e.** Which (if any) missing stakeholder viewpoints would strengthen your team's assessment of this system's potential impact on human rights and fundamental freedoms?

...................................................

## Component 3: Engagement Objective

**a.** Why are you engaging stakeholders?

...................................................

**b.** What do you envision the purpose and the expected outcomes of engagement activities to be?

...................................................

**c.** Ideally, how would stakeholders be able to influence the engagement process and the outcomes?

...................................................

**d.** What engagement objective do you believe would be appropriate for this project considering challenges or limitations to assessments related to positionality, and proportionality to the project's potential degree of impact?

...................................................

**e.** Considering answers to the above questions, what is your established engagement objective?

...................................................

## Component 4: SIA Engagement Method

**a.** What resources are available and what constraints will limit potential approaches?

......................................................

**b.** Which methods meet your team's engagement objective?

......................................................

**c.** What accessibility requirements might stakeholders have?

......................................................

**d.** Will online or in-person methods (or a combination of both) be most appropriate to engage salient stakeholders?

......................................................

**e.** Considering the above questions, what is your established engagement method for the SIA?

......................................................

**f.** How will your team make sure that this chosen method accommodates different types of stakeholders?

......................................................

**g.** How will your team ensure that, where appropriate, the PS Report used to conduct the SIA is accessible to stakeholders?

......................................................

**h.** How will your team ensure that your engagement method feeds useful information to your stakeholder impact assessment? Consider what feedback mechanisms will be in place.

......................................................

AI Sustainability in Practice Part One:
Foundations for Sustainable AI Projects

# Activities

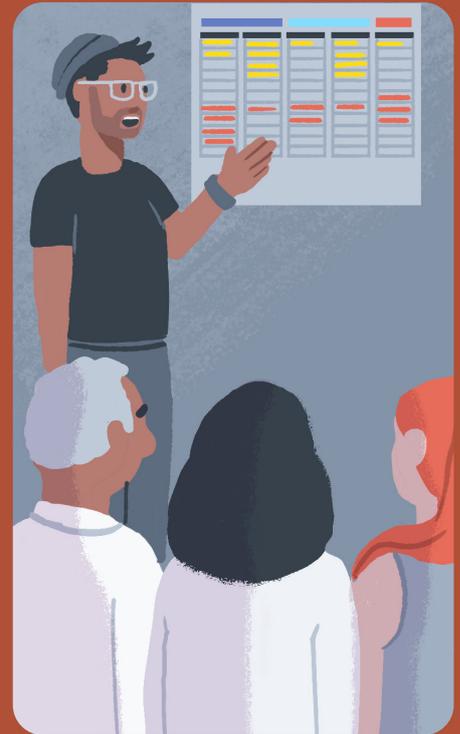

# Activities Overview

In the previous sections of this workbook, we have presented an introduction to the core concepts of AI Sustainability. In this section we provide concrete tools for applying these concepts in practice. Activities related to AI Sustainability in Practice Part One will help participants engage with the SUM Values and practice establishing the foundations for sustainable AI projects. This workshop is to preface the AI Sustainability in Practice Part Two workshop, where participants will practice implementing governance actions related to the principle of Sustainability.

We offer a collaborative workshop format for team learning and discussion about the concepts and activities presented in the workbook. To run this workshop with your team, you will need to access the resources provided in the link below. This includes a Miro board with case studies and activities to work through.

Workshop resources for **AI Sustainability in Practice Part One:** [turing.ac.uk/aieg-2-activities](turing.ac.uk/aieg-2-activities)

## A Note on Activity Case Studies

Case studies within the Activities sections of the AI Ethics and Governance in Practice workbook series offer only basic information to guide reflective and deliberative activities. If activity participants find that they do not have sufficient information to address an issue that arises during deliberation, they should try to come up with something reasonable that fits the context of their case study.

### Note for Facilitators

In this section, you will find the participant and facilitator instructions required for delivering activities corresponding to this workbook. Where appropriate, we have included Considerations to help you navigate some of the more challenging activities.

Activities presented in this workbook can be combined to put together a capacity-building workshop or serve as stand-alone resources. Each activity corresponds to a section within the Key Concepts in this workbook. Some activities have pre-requisites, which are detailed on the following page.

**Relating to Values**

Demystify AI ethics by building a common vocabulary of SUM Values grounded on your group's personal and collective relationship to each of them.

**Corresponding Sections**
→ Introduction to Sustainability: SUM Values (page 10)
→ The SUM Values in Focus (page 15)

---

**Stakeholder Analysis**

Practise identifying vulnerable stakeholders by anticipating specific project's impacts on individuals and communities.

**Corresponding Sections**
→ Preliminary Project Scoping and Stakeholder Analysis (page 24)

---

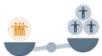

**Stakeholder Prioritisation**

Practise evaluating stakeholder prioritisation.

**Corresponding Sections**
→ Preliminary Project Scoping and Stakeholder Analysis (page 24)

**Pre-Requisites**
↗ Activity: Stakeholder Analysis (page 59)

---

**Positionality Reflection**

Practise reflecting on your team's positionality with respect to case-specific stakeholders.

**Corresponding Sections**
→ Determining a proportionate approach to stakeholder involvement (page 26)
→ Positionality Reflection (page 28)

---

**Establishing an Engagement Objective**

Practise tailoring stakeholder participation goals to the needs of specific projects.

**Corresponding Sections**
→ Determining a proportionate approach to stakeholder involvement (page 26)
→ Determining Stakeholder Engagement Objectives for Stakeholder Impact Assessmnets (page 30)
→ Determining Stakeholder Engagement Methods for Stakeholder Impact Assessmnets (page 34)

**Pre-Requisites**
↗ Activity: Stakeholder Analysis (page 59)
↗ Activity: Stakeholder Prioritisation (page 67)
↗ Activit: Positionality Reflection (page 69)

# Interactive Case Study: AI in Urban Planning

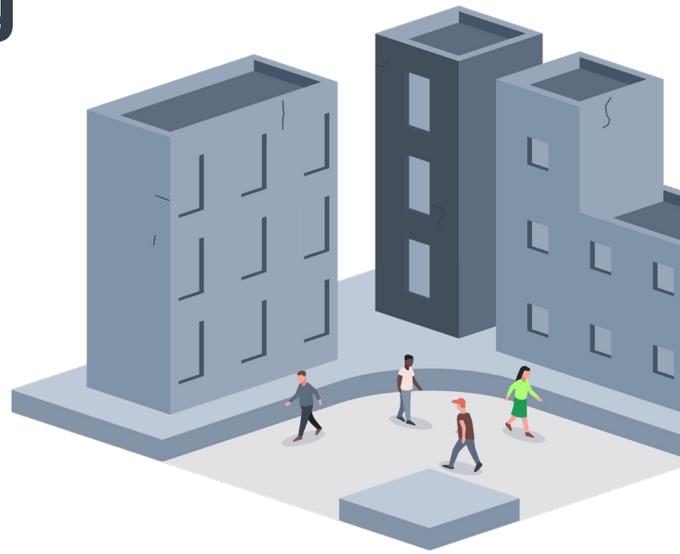

Your team is a local planning authority within a borough facing a housing crisis. The local poverty rate is higher than the national average and residents complain of sub-optimal living conditions.

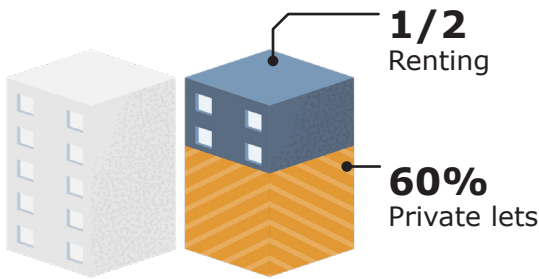

**1/2**
Renting

**60%**
Private lets

Around half of your residents are renters, 60% of whom live in private lets. The private letting sector is becoming increasingly unaffordable.

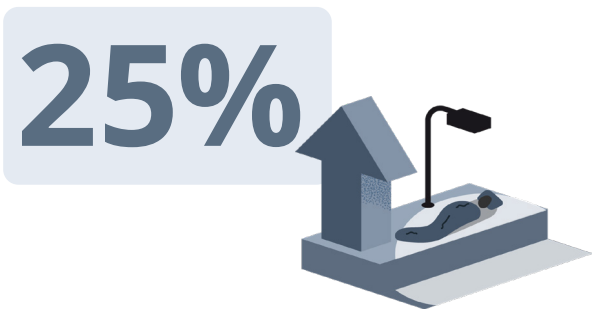

**25%**

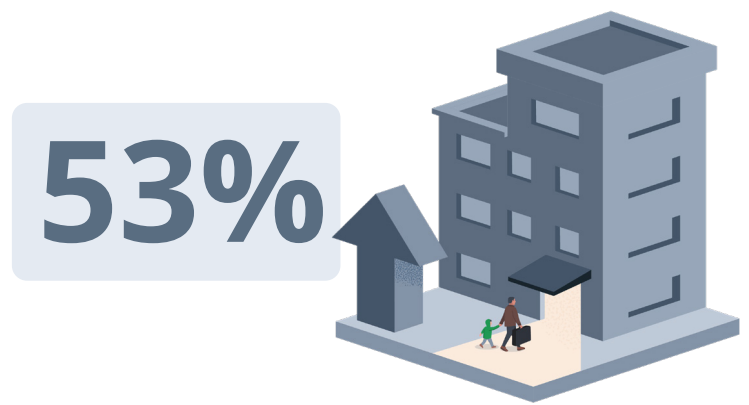

**53%**

The number of homeless applications has **risen by 25% in the past three years.**

**The number of households in temporary accommodation has risen by 53%,** with an unprecedented number of applications submitted since 2020.

A recent council investigation found that **terminated private tenancy leases** are the **single greatest cause of homelessness in the borough.**

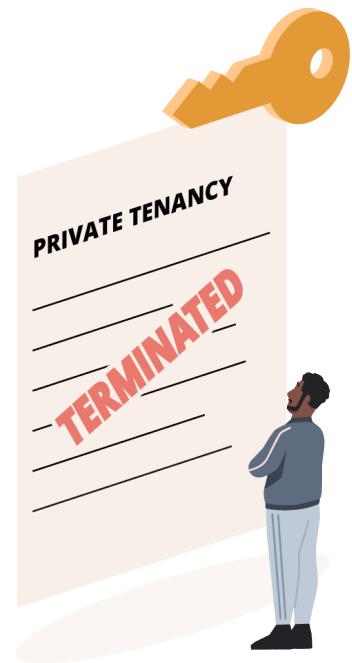

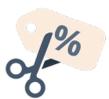

**10,000**
New homes

**50%**
Affordable homes

Your council has established a 10-year housing plan set out to deliver 10,000 homes, 50% of which will be affordable. The objective of this plan is to improve the living standards of residents by developing as many high-quality affordable homes as possible over the next ten years.

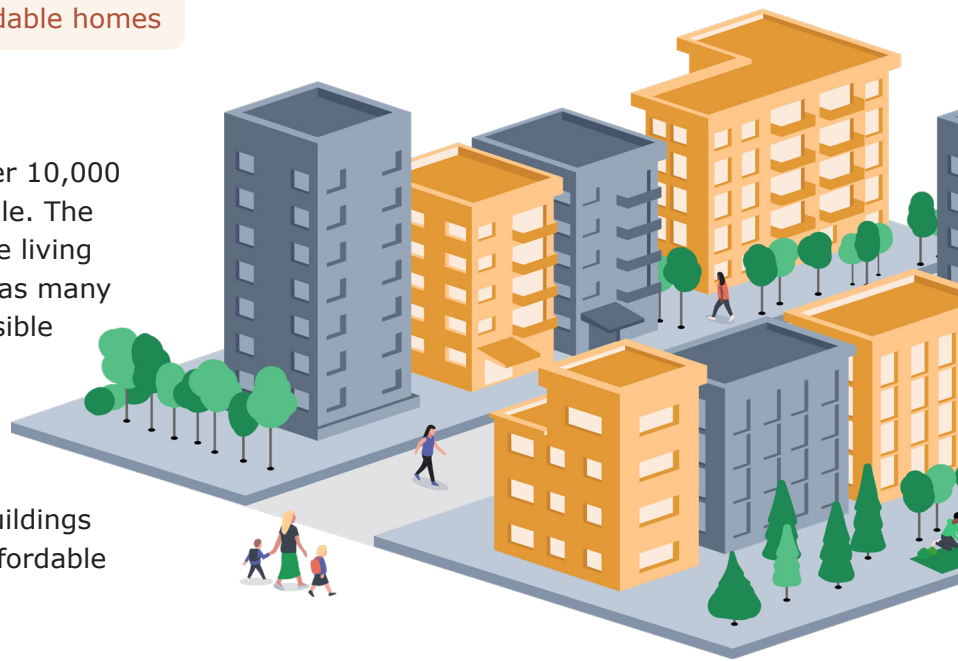

The council has offered to subsidise new residential buildings that deliver at least 50% affordable housing.

To support housing developments, your team will need to **expand the list of sites permitted for planning applications.** Achieving your target would mean **doubling** the number of local homes. Your team will need to review a much higher volume of planning applications, which may not be achievable through your current process.

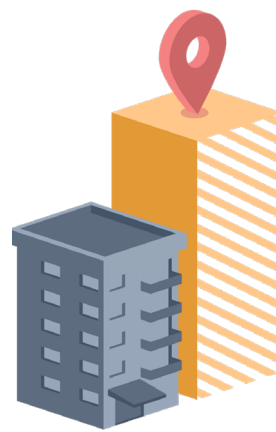

# Model Proposal

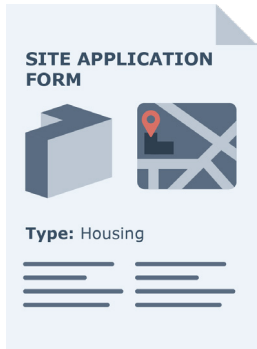

## Current Method

Your current method for allocating new development sites can take up to ten months to complete and considers **a limited number of sites proposed by developers, landowners, and estate agents.** These sites are manually reviewed by your team to ensure they meet policy standards (i.e. sites' ability to provide basic amenities) and are suitable for development in practice.

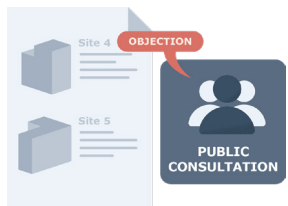

**Sites that pass your review process are taken forward for a public consultation.** This gives residents the opportunity to object to certain sites being open for planning applications. Your team considers public input to help determine which site proposals are accepted.

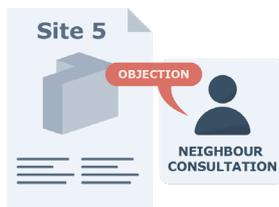

**Accepted sites are made available for planning applications.** Applications are detailed development proposals demanding in-depth review. **Your team manually reviews individual applications in a process that includes a second tier of consultations with neighbours of specific sites.**

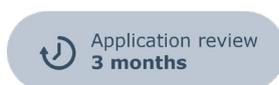

Granting planning permissions can take up to three months per application.

### Proposed Method

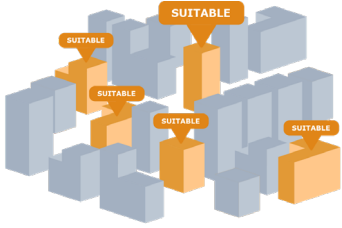

Your council has suggested you automate this process by using an machine learning (ML) model to automatically review every site in the local area, classifying them as suitable or unsuitable for housing development.[59] This approach would allow your team to scale-up the number of sites considered for development. Whereas your current method captures a number of submitted proposals, **the model would capture all local sites.** This model would consider sites that are outside the reach of your current method, such as council owned buildings that could be repurposed, and private parcels that could accept purchase offers.

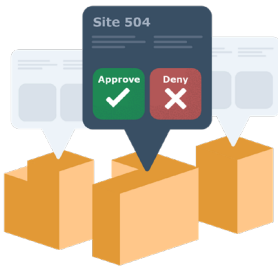

Sites categorised as suitable would be reviewed by your team. **Those that pass this review process would be brought forward for a three month public consultation** which your team would consider when accepting a final list of sites for development. **Accepted sites would be made public in a digital map and approved for development, forgoing the additional three-month application review process.**

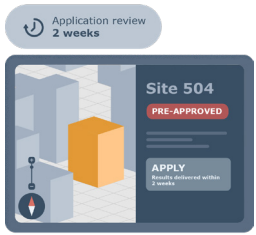

**The proposed method would remove neighbour consultations from the application review process.** By removing time-consuming steps, your team would be able to verify applications' compliance with building design standards and grant approvals or request adjustments within two working weeks.

# Your Task

The council has conducted desk research about the context of the project and provided your team with examples of public responses to similar projects conducted by other local authorities in the past. **Your team is to draw on this information to begin your Stakeholder Engagement Process and evaluate the possible impacts and ethical permissibility of the proposed system.**

# Desk Research Materials

The following materials are reactions to other councils using this model in the past.

## A Wave of Affordable Homes Is Coming and We've Got Computers to Thank for That.

**MIRJAM NILSSON**
OPINION

A strong example of this has been at our local council, where a new AI system has helped discover new locations optimal for housing development, and streamline decision-making. Paired with subsidies for developments that deliver at least 50% affordable homes, the council's approach is demonstrating genuine **care** for those in desperate need of housing.

The developments in the former Local Gardens, the old commercial centre, and on top of thirty high street shops are just a few on the list of subsidised buildings that will attract disposable income into the area, and with that, an array of shops and services that will provide employment to lower income residents.

**SUM Values Referenced:**

**Care** for the wellbeing of each and all

**Local Environmentalist**
@Biospheric_Care

It is frustrating that the few green spaces in London - where families and friends can **connect** - are not being cared for by local councils! #**Care**ForTheEnvironment #**Connect**ThroughNature

1:21 PM • Feb 25, 2021

**17** Retweets   **2** Quote Tweets   **359** Likes

**SUM Values Referenced:**

**Care** for the wellbeing of each and all

**Connect** with each other sincerely, openly, and inclusively

**LOCAL NEWS TODAY**

# First 'Al Green Light' Council-Owned Housing Development to Begin Today

The construction of a 200 unit mixed income housing development is to start today at a local public area. Today's construction comes in record time, as it was pre-approved by a council algorithm and submitted for a "basic check" just three weeks ago. This is the second council project to be built under the new local plan. What is seen by many as an innovation that will speed up the delivery of urgently needed housing, is critiqued by others for a lack of local **autonomy** over what sites are allowed for construction.

> "We will solve local homelessness with the help of automation. Our local plan we will provide as many possible homes to the most possible people, as soon as possible."
>
> - Local Council Spokesperson

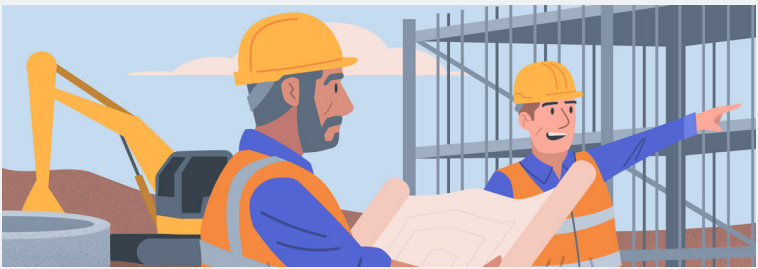

**SUM Values Referenced:**

**Respect** the dignity of individual persons

**House, P.J.**
@ProtectJustice

Finally, AI for social good! New algorithm helping determine locations for affordable housing, **protecting** historically marginalised communities. Still lots to be done, but definitely a step in the right direction. #AIForJustice #**Protect**WithHousing

1:21 PM • Feb 25, 2021

**29** Retweets   **15** Quote Tweets   **402** Likes

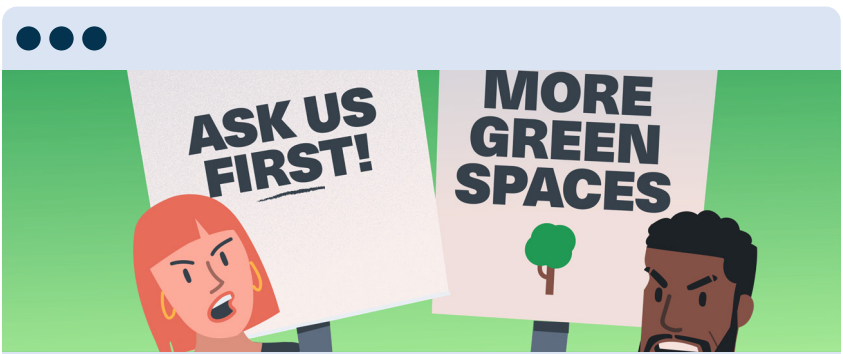

## COUNCIL DEVELOPMENT PROTEST

A local council algorithm has selected the Local Gardens as the most suitable place to build flats. Whilst the development will provide 100 affordable homes at a time where many are cannot afford local rent prices, neighbour's of the site protested at the construction site today, expressing their concerns with not being **connected** to decision-making in the local plan.

> Many of us were not aware of the council's plan to take away our garden, and now we are not entitled to refuse. This is not consensual. No **respect** for residents!

- Local Resident

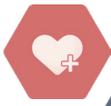 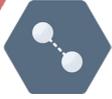

# Relating to Values

## Objective

The purpose of this activity is to build a common vocabulary and understanding of SUM Values by reflecting on your team's personal and collective relationship to each of these. This reflection will help the team think more intuitively about the moral scope and the ethical impacts of AI projects.

## Team Instructions

**1.** Your team will divide into groups, each is assigned a SUM Value.

**2.** In each group, have a discussion about how this value might play a part in your lives. Consider the questions:

- Is this value important to you? If so, why?

- Do you strive to practice this value? If so, in what contexts?

- Have you worked in a team or organisation where one or more of these values was an active part of the culture? If so, how did this team or organisation demonstrate this value? Can you think of an example?

- Have you worked in a project where this value was actively considered? If so, how might this project have benefited from this consideration?

**3.** Use sticky notes to write your answers and paste these on the **Values Map**.

**4.** Reconvene and share your reflections, discussing as a team and adding new reflections to the **Values Map**.

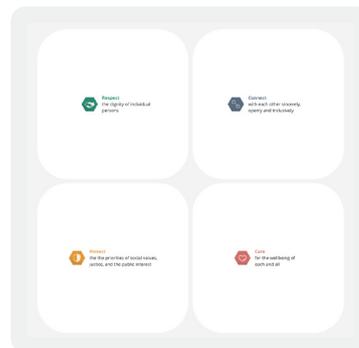

**Values Map**

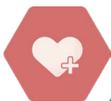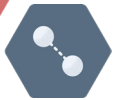

30 mins | Facilitator Instructions

# Relating to Values

## Part One: Group Reflections  15 mins

1. Let the team know that you will now engage in your first activity.

2. Give them time to read the objective and team instructions for this activity, asking them if they have any questions.

3. Divide the team in groups, assigning a SUM Value to each group.
   - If delivering digitally, split each group into a breakout room.

4. Ask for a volunteer note-taker from each group. Note-takers will also be tasked with reporting back when groups reconvene.

5. Invite each group to discuss how the assigned value might play a part in their lives.

## Part Two: Sharing and Discussing  15 mins

1. Let them know that each team will have some minutes to share their reflections.

2. After each group shares, ask the team to share anything they might like to add to the **Values Map**.
   - Ask them to write their responses in sticky notes and place them on the map.

3. Let the team know that they can refer to this map throughout the workshop to refresh their memory on what the SUM Values mean to them.

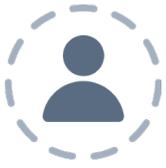

# Stakeholder Analysis

## Objective
The purpose of this activity is to practice identifying project impacts on specific individuals and communities.

## Overview
You will be split into groups for this activity. Each group will be assigned three stakeholder profiles to analyse.

- Your facilitator will ask for a volunteer note-taker for each group.
- Feel free to use the example profile of 'Hayley' as reference throughout the activity.

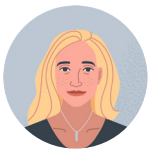

## Hayley
**40** **She/her**

### Profile
Hayley is a 40-year-old white British woman on the housing register. She lives in an overcrowded flat with her family of five, she is waiting for a bigger home ideally in proximity to affordable childcare and a specialist school for her son who has autism spectrum disorder (ASD).

### Goals and Aspirations
Hayley and her husband's living situation has been overcrowded since the birth of their third child a year ago, and greatly worsened when they started working from home. The lack of space has been extremely challenging for their eldest son in particular, with ASD. Their current flat is in close proximity to their eldest son's specialist school, and affordable childcare for their two youngest. They are hoping to move to a bigger home that provides this level of access as soon as possible.

### Stakeholder Qualities Chart

**Protected Characteristics**

Woman, legal guardian of person with disabilities, legal guardian of children.

**Vulnerability Factors**

Person in housing register, person with special accommodation needs.

### Stakeholder Groups

Children, disabled communities, people in housing register, and people with special accommodation needs.

### Possible Impacts

**Potential Harms**

The system could identify sites without access to specialist education for her son, neglecting equity and the value of Protect.

**Potential Benefits**

The system could speed up housing register wait times, allowing her to obtain a house sooner, prioritising public interest and the value of Protect.

## Part One: Identifying Stakeholders ⏱ 15 mins

1. As a team, read over your assigned **Stakeholder Profiles**.

2. Pick one of the profiles and use the prompts within the **Stakeholder Qualities Chart** to write down protected characteristics and vulnerability factors within this profile. Repeat this for all of your group's assigned profiles.

   - **Protected Characteristics** include age, disability, gender reassignment, marriage and civil partnership, pregnancy and maternity, race (including ethnicity, colour, nationality), religion or belief, sex, sexual orientation.[60]

   - **Vulnerability Factors** are factors and specific circumstances that fall outside of protected characteristics which may expose them to being jeopardised by a project.

3. As a group, discuss how each stakeholder's protected and contextual characteristics might align with relevant stakeholder groups. For each of these characteristics, consider the question:

   - What affected group or groups of stakeholders might this characteristic or quality represent?

4. Use the **Stakeholder Groups** column within the **Stakeholder Profiles** to list social groups represented by these qualities (i.e. houseless people, business owners, children).

## Part Two: Scoping Potential Stakeholder Impacts ⏱ 10 mins

1. Answer the questions in the **Possible Impacts** boxes within your **Stakeholder Profiles**. Write your group's answers on sticky notes, placing them within their respective boxes.

2. Having identified potential harms and benefits or each stakeholder, have a group deliberation about how these relate to one or more SUM Values. Consider:

   - How do these harms or benefits support or neglect one or more SUM Values?

## Part Three: Group Discussion ⏱ 15 mins

1. Your facilitator will ask the team to reconvene.

2. Each group's note-taker will report back to the team, sharing:

   - How the group found each stakeholder to be impacted differently.

   - What stakeholders the team identified as salient.

3. Have a group discussion about this activity.

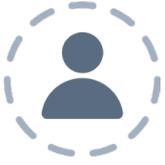

# Stakeholder Analysis

1. Read out the context for this activity, asking the team if they have any questions:

> **Activity Context**
> Our council has shared recent information about residents and council employees that have participated in interviews for a council research project. Most of the identities and experiences contained in the sample of personas provided represent overarching trends within individuals and communities that could be impacted by the system we are assessing. The team will use the sample group and research on other councils who have used this model as a starting point for analysing project stakeholders.

2. Give the team a moment to read over the instructions.

3. Next, give them some minutes to review the **Case Study** and **Desk Research Materials**.

4. Take a moment to explain the activity, letting the team know that each group will be tasked with using their assigned stakeholder profiles to conduct a stakeholder analysis.

   - Let them know that they can use the stakeholder profile 'Hayley' as reference throughout the activity and that you will be available to answer questions at any point.

5. Split the team into groups (each assigned three stakeholder personas). For example:

   - **Group 1**: George, Tom, Katherine
   - **Group 2**: Terry, Jamie, Mia
   - **Group 3**: Alex, Ali, Nick

6. Make yourself available for each of the groups, using the considerations section of this activity to answer questions and aid the groups.

 **Stakeholder Analysis**

Here are some additional details that may help facilitate group discussion:

- **Protected Characteristics**
  In the 2010 UK Equality Act, protected classes include age, gender reassignment, being married or in a civil partnership, being pregnant or on maternity leave, disability, race including colour, nationality, ethnic or national origin, religion or belief, sex, and sexual orientation. The European Convention on Human Rights, which forms the basis of the UK's 1998 Human Rights Act, includes as protected characteristics 'sex, race, colour, language, religion, political or other opinion, national or social origin, association with a national minority, property, birth or other status. For a more detailed description, see the Equality and Human Rights Commission's webpage on protected characteristics.

- **Vulnerability Factors**
  For the purposes of this activity, vulnerability factors can include individuals in need of housing (with increased vulnerability for homeless people, those in unsanitary or overcrowded housing, those in need to live near specialist medical or educational facilities), those in economically disadvantaged positions, those whose goals and aspirations may be compromised by development (i.e. residents who use sites that might be repurposed).

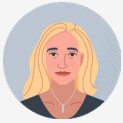

### Hayley  `Impacted Stakeholder`

**Protected Characteristics**
Woman, legal guardian of person with disabilities, legal guardian of children.

**Potential Harms**
The system could identify sites without access to specialist education for her son, neglecting equity and the value of Protect.

**Vulnerability Factors**
Person in housing register, person with special accommodation needs.

**Potential Benefits**
The system could speed up housing register wait times, allowing her to obtain a house sooner, prioritising public interest and the value of Protect.

**Social Groups**
Children, disabled communities, people in housing register, and people with special accommodation needs.

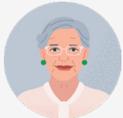

### Katherine  `Impacted Stakeholder`

**Protected Characteristics**
Elderly person and disabled person.

**Potential Harms**
Without proper care for the quality of the model's outcomes, it could identify sites without appropriate access to transport and leisure facilities.

**Vulnerability Factors**
Local community member in housing register – needs to live near leisure facilities and transport, needs accessible accommodation.

**Potential Benefits**
With proper care for the quality of the model's outcomes, it could provide Katherine with a home suitable for her mobility needs which also supports her lifestyle through access to transport and leisure facilities.

**Social Groups**
Elderly communities, disabled communities, and local residents.

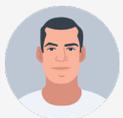

### Michael  `Project Team Member`

**Potential Benefits**
The successful delivery of an AI project that supports the housing delivery plan could have a positive impact on his career goals.

**Potential Harms**
Risks contained within the deployment of this model could reduce the quality of life and harm residents, which could in turn have an impact on him.

**Vulnerability Factors**
Team member of planning authority.

**Protected Characteristics**
None

**Social Groups**
Local residents.

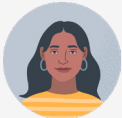

**Mia** <span style="color:blue">**Project Team Member**</span>

**Potential Harms**
Risks contained within the deployment of this model could reduce the quality of life and harm residents, which could in turn have an impact on her.

**Protected Characteristics**
Person from ethnic minority background (Indian).

**Potential Benefits**
The successful delivery of an AI project that supports the housing delivery plan could have a positive impact on her career goals.

**Vulnerability Factors**
Team member of prospective planning authority.

**Social Groups**
Indian communities.

---

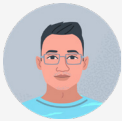

**Alex** <span style="color:blue">**Project Team Member**</span>

**Potential Harms**
Developing and deploying the model without proper involvement of residents and planning authorities could hamper consideration of their viewpoints and authority in urban planning processes.

**Potential Benefits**
Developing and deploying the model with thorough community involvement and the involvement of the planning authority team could respect their knowledge and authority while delivering greater quantities of homes at speed.

**Protected Characteristics**
Person from ethnic minority background (Chinese).

**Vulnerability Factors**
Team member of planning authority.

**Social Groups**
Chinese communities.

---

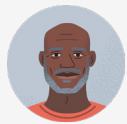

**George** <span style="color:red">**Impacted Stakeholder**</span>

**Potential Harms**
The site where his business is located could be identified as suitable, putting his lease at risk.

**Potential Benefits**
The model will attract more residents which could be potential customers for George's business.

**Protected Characteristics**
Person from ethnic minority background (black).

**Vulnerability Factors**
Local business owner.

**Social Groups**
Black communities and local business owners.

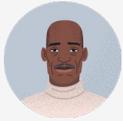

### Tom  *Impacted Stakeholder*

**Potential Harms**
None.

**Potential Benefits**
The model will support Tom's goal to sell some of his real estate portfolio by promoting development in the area.

**Protected Characteristics**
Immigrant person from ethnic minority background (black).

**Vulnerability Factors**
None.

**Social Groups**
Immigrants and black communities.

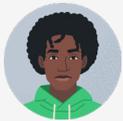

### Ali  *Impacted Stakeholder*

**Potential Harms**
His garden could be categorised as a suitable site, putting it at risk his involvement in the community as well as the local character (composed in part by community spaces).

**Potential Benefits**
None.

**Protected Characteristics**
Immigrant person from ethnic minority background (mixed race).

**Vulnerability Factors**
Local community member, child.

**Social Groups**
Mixed race communities, immigrants, local residents, and children.

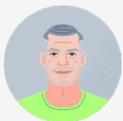

### Nick  *Impacted Stakeholder*

**Potential Harms**
The model could identify sites in areas that are unsafe for transgender people.

**Potential Benefits**
None.

**Protected Characteristics**
Person from gender minority background (transgender man).

**Vulnerability Factors**
Local community member, homeless individual.

**Social Groups**
Transgender communities, homeless communities, and local residents.

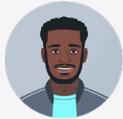

**Terry**  *Impacted Stakeholder*

**Protected Characteristics**
Person from ethnic minority background (black).

**Potential Harms**
The target variable for suitability indicates sites being available for mixed income housing, which may cause property value in the area to go up, raising rent prices further.

**Vulnerability Factors**
Local community member.

**Social Groups**
Black communities and local residents.

**Potential Benefits**
None.

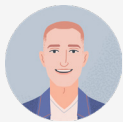

**Jamie**  *Impacted Stakeholder*

**Protected Characteristics**
Person from sexual minority background in a homosexual marriage.

**Potential Harms**
None.

**Vulnerability Factors**
Local resident.

**Potential Benefits**
The model will make available real estate for sale in the area, supporting his objective of purchasing a home.

**Social Groups**
LGBTQ+ communities.

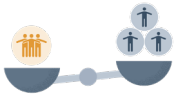

# Stakeholder Prioritisation

## Objective

The purpose of this activity is to practice evaluating stakeholder prioritisation. Through engaging with a hypothetical case study, participants will identify which stakeholders are likely to be most impacted, vulnerable, and those that currently have least influence over the project.

## Team Instructions

1. You will be split into groups for this activity. One person from each group is to volunteer to take notes.

2. With your group, read over the stakeholder personas and the desk research provided. Have a discussion considering the following questions:

   • Can you think of other harmful or beneficial impacts that this project may have on each stakeholder?

   • How might these personas be differently impacts by this project?

   • Are there any relevant power relations between these personas? Consider their relative advantages and disadvantages, and what personas may have influence over the project and its outcomes.

3. Building on your discussion, consider the following questions, discussing what stakeholder persona (or multiple personas) you think needs to be prioritised in this project:

   • What persona(s) is(are) likely to be most significantly impacted by this project? Considering:

     – What persona(s) has(have) the greatest needs in relation to potential benefits of the system?

     – What persona(s) is(are) most vulnerable to being severely harmed by this project?

     – What persona(s) is(are) likely to have limited influence on this project?

4. Having had your discussion, write down salient stakeholders in sticky notes, placing them in the **Prioritised Stakeholders** section.

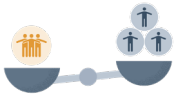

# Stakeholder Prioritisation

1. Give the team a moment to read over the instructions.

2. Take a moment to explain the activity, letting the team know that each group will be tasked with using their assigned stakeholder profiles to identify which stakeholders are likely to be most impacted, vulnerable, and those that currently have least influence over the project.

3. Split the team into groups. If this activity is conducted along with the Stakeholder Analysis activity, keep the same groups.

4. Make yourself available for each of the groups.

5. When enough time has passed, ask the team to reconvene.

6. Give each group's note-taker a few minutes to share back to the greater team, considering:

   - How the group found each stakeholder to be impacted differently

   - What stakeholders the team identified as needing to be prioritised

7. Use the remaining time for the team to discuss the activity.

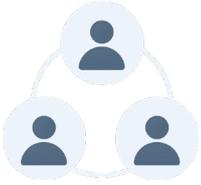

# Positionality Reflection

## Objective

Practise reflecting on team's positionality with respect to impacted stakeholders.

## Overview

In this activity, the group will be asked to consider the profiles of the members of the local authority team introduced in the case study when conducting a positionality reflection.

When considering team positionality, it is important to understand and contextualise why certain characteristics may serve as strengths when considering ethical permissibility, while others, while preserving the possibility for valuable contributions, may present limitations and the risk of unconscious bias. This comes from an understanding of individual experiences as sources of valuable knowledge, as well as a recognition of historic inequalities that contribute to current social contexts, which produce asymmetrical distributions of agency, access, and life chances.

The group will reflect on the team positionality in relation to the project stakeholders that have been identified as prioritised stakeholders in the previous activity.

## Team Instructions

1. In the **Prioritised Stakeholders** section, select the personas from the list above whom you have identified as Prioritised stakeholders in the previous activity.

2. Individually read through the **Prioritised Stakeholder** section, the **Team Member Personas**, and review the **Case Study** section if needed.

3. Next, your team will be divided into groups.

4. As a group, read through the **Positionality Matrix** section, considering how each persona in the **Team Member Personas** section may respond to the questions provided. Use sticky notes to write on their respective section of the matrix.

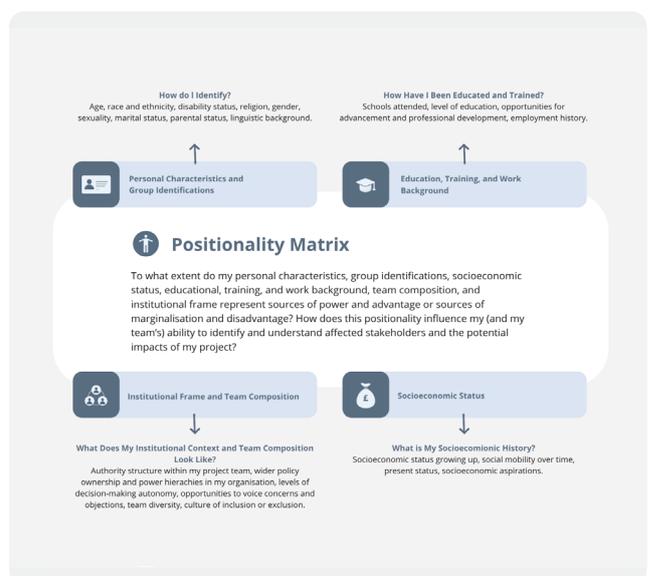

**Positionality Matrix**

- **Note:** Write the persona's name on sticky notes.

**5.** As a group, discuss the collective positionality of the project team, answering the questions below:

- How does their team positionality relate to that of prioritised stakeholders (and their dependants)?

- Are there ways that their positionality as a team could limit their perspective when assessing this AI system?

- Are there ways that their position as a team could strengthen their perspective when assessing this AI system? Consider overlapping identities and experience.

- What (if any) prioritised stakeholder viewpoints are currently missing in their team composition?

**6.** As you discuss, your facilitator will write strengths within notes on the **Hot Air** section. These notes represent factors that strengthen (or "elevate") the team's analysis of this project.

They will write any limitations (factors that "weigh your team down") to the project team's positionality within notes, placing them in the **Sandbags** section. These notes represent current challenges or limitations to assessing this project given the project team's positionality.

Lastly, your co-facilitator will write missing stakeholder viewpoints within notes, placing them in the **Missing Viewpoints** section. These notes represent missing viewpoints that would further strengthen the project team's assessment of this project, which the team will aim to include through stakeholder engagement.

**7.** Reconvene as a team and discuss your reflection.

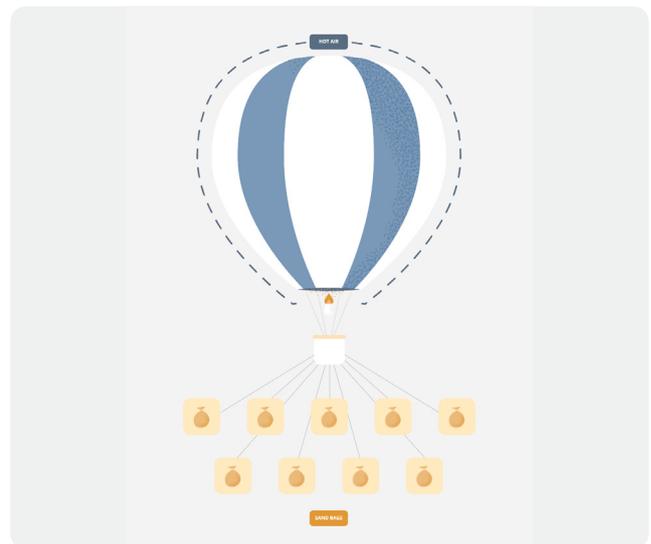

**Hot Air Ballon**

*This activity was adapted from The Feminist Design Tool, created by Feminist Internet and Josie Young.*[6]

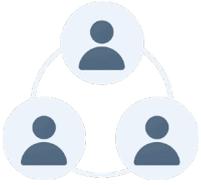

# Positionality Reflection

1. Give the team ten minutes to fill in the questions within the **Positionality Matrix**.

2. Divide the team into groups.

3. Facilitator and co-facilitators: Guide a group reflection by asking the group the questions in the participant instructions. Use the considerations section of this activity to aid in this activity and answer any questions.

- **Co-facilitator:** as the team discusses, write down relevant notes in the **Sand Bags**, **Hot Air**, and **Missing Viewpoints** sections.

The concept of 'rank' refers to the system of valuing people differently depending on certain social memberships. Different characteristics within single individuals may belong to groups with power and advantage or marginalised and disadvantaged social groups. These belongings are correlated with levels of access within asymmetrical distribution, as well as the likelihood of experiencing specific forms of privilege and oppression.[62]

| Rank Category | Groups with Power and Advantage | Marginalised and Disadvantaged Groups |
|---|---|---|
| Age | Young and middle-aged adults | Children and the elderly |
| Disability status | Non-disabled people | People with cognitive, intellectual, sensory, physical, and/or psychiatric disabilities |
| Religion | Christian and secular | Muslims, Jews, Hindus, Buddhists, and other minority religions |
| Ethnic and racial identity | White people | Black, Asian, South Asian, Latino, African, Middle Eastern, multiracial, and other ethnic or racial minorities |
| National origin | British people | Non-British people |
| Gender | Cisgender men | Women, Transgender, Non-binary, and other minority genders |
| Linguistic background | English speakers | Non-English speakers |

Privilege stems from belonging to social groups which have historically been valued and whose interests have been, either intentionally or unintentionally, considered when establishing cultural and institutional practices, the design of technology as well as a variety of contextual mechanisms that produce asymmetrical distribution. Although power and advantage may look like having access to basic rights and being treated with dignity, the understanding that these groups are privileged is rooted in the fact that such a favourable level of access is not equally distributed among social groups.

Marginalised groups refer not necessarily to those whose membership comprises a smaller portion of a population than other related groups, but rather to those possessing less access to privilege and power. The oppression of marginalised groups stems from them being historically undervalued. Likewise, the interests of marginalised groups are often not considered or prioritised in the decision-making processes and mechanism that shape their lives and their opportunities. They both risk being overlooked while simultaneously possessing heightened vulnerabilities.

Everyday constructs produce mechanisms of asymmetry, for example:

- Elderly individuals are disproportionately selected for workplace redundancy.

- Most homes are not built with essential features required by those with mobility needs.

- Black individuals (including children) are more likely to be stopped and searched than their white counterparts.

- Transgender individuals face disproportionate targeted street-based and online violence as opposed to their cisgender counterparts.

- Women are more likely to occupy jobs associated with less status and pay in AI and data science.

Beyond personal characteristics and group identifications, qualities such as socioeconomic status and education, training and work background are also subject to asymmetrical power access. Consider the potentially different experiences, aspirations, power, and access of individuals in these different categories:

| Positions of Power and Advantage Pertaining to Socioeconomic Status | Positions of Marginalisation and Disadvantage Pertaining to Socioeconomic Status |
|---|---|
| Raised in upper- and middle-class households | Raised in working-class or economically deprived households |
| Who have developed social mobility into upper- or middle-class status | Who have either remained working-class or economically deprived, or whose economic status has diminished |
| Who are currently upper- or middle-class | Who are working-class or economically deprived |

## Education, Training and Work Background

| Positions of Power and Advantage Pertaining to Education, Training, and Work Background | Positions of Marginalisation and Disadvantage Pertaining to Education, Training, and Work Background |
|---|---|
| Attended private, elite, and/or well-funded schools with adequate resources. | Attended underfunded public schools with a lack of adequate resources. |
| Are or have been present in contexts with available opportunities for advancement and professional development, and/or have had the ability to pursue these opportunities. | Are or have been present in contexts were opportunities for advancement and professional advancement are rare or non-existent, or who aren't or have not been able to pursue opportunities due to contributing factors related to their identity characteristics and/or socioeconomic status. |
| Have an employment history that provides relatively secure career prospects and advancement. | Have an employment history that does not provide relative security for the future of their career and advancement. |

The reality of asymmetries presented by this array of factors are deeply ingrained in our social conditioning and may be seen as normal or the default. It is therefore easy to not be aware of inequalities when they are not part of our lived experience. It takes intentional work to consider the interests of marginalised and disadvantaged groups which we are not part of. These standpoints are therefore crucial when it comes to assessing ethical permissibility because they can provide knowledge and insight grounded in lived experiences, which are likely to otherwise be overlooked.

Although having a team composition that reflects a diversity of marginalised standpoints is a good starting point for considering the interests of marginalised individuals, your team's institutional frame, including authority structures, power hierarchies, opportunities to voice concerns, and objections and cultures of either inclusion or exclusion will either facilitate or hinder processes of inclusion and your team's ability to integrate these standpoints. For example, the decision to engage stakeholders may be sidestepped by teams where inclusion is not a salient value. A team with members representing a diversity of marginalised standpoints, but who are only present in lower ends of power hierarchies and are not supported in practicing decision-making autonomy, is unlikely to meaningfully integrate their insights and knowledge.

### How Does Your Team Positionality Relate to that of Project Stakeholders?

This question relates directly to institutional frame and team composition. It challenges teams to compare their team composition to that of project stakeholders. Some questions to consider in facilitating include: Are there overlapping identities? Do individuals possessing these overlapping identities have the power to significantly influence decision-making? Are there any obvious asymmetries of power and access between the team and stakeholders?

### Are There Ways that Your Position as A Team Could Limit Your Perspective when Assessing This AI System?

This question challenges teams to mindmap potential assumptions made about stakeholders based on a lack of overlapping identities and experiences. Team members are encouraged to consider elements of stakeholders' identities and experiences that they might be unfamiliar with, reflecting on whether there are needs or vulnerabilities that could be best identified directly.

Potential reflection questions include:

- Are there areas where the team might incorrectly assume that value will be added to stakeholders?

- Are there considerations that the team could leave out or deem as unimportant based on a lack of overlapping identities or experiences?

### Missing Stakeholder Standpoints

The specific stakeholder standpoints identified for this part of the activity will differ from team to team. This activity will be supported by the previous questions. The challenge posed to the team is to explicitly write down what identity characteristics, socioeconomic backgrounds, education, training, and work backgrounds are present within identified stakeholders, but missing from the team composition. These standpoints are likely to provide valuable insights that might otherwise be omitted when assessing ethical permissibility.

### Further Support

The subject of positionality and inequality can be challenging for teams and facilitators. Developing skills in this area and moving closer to supporting equal power and advantage takes effort and begins with an explicit recognition of our own areas of privilege and marginalisation.

The following resources may support facilitators in further preparing for this activity in advance:

*Understanding Oppression: Strategies in Power and Privilege*
by Leticia Nieto and Margot F. Boyer

*Understanding Oppression Part 2: Skill Sets for Targets*
by Leticia Nieto and Margot F. Boyer

*Understanding Oppression Part 3: Skill Sets for Agents*
by Leticia Nieto and Margot F. Boyer

*Diversity, Equity, and Inclusion Practice Guide*
by 501 Commons

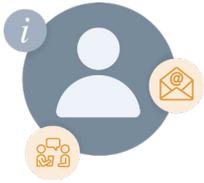

# Establishing an Engagement Objective

## Objective
Practise tailoring stakeholder participation goals to the needs of specific projects.

## Activity Context
Having analysed project stakeholders and reflected on team positionality, you will now address potential challenges and limitations in assessing this AI system by establishing an engagement objective that enables the appropriate degree of stakeholder involvement in Stakeholder Impact Assessments (SIAs).

## Team Instructions

1. Your team will be split into groups. In your groups, consider the following engagement objectives:

   - **Inform:** Make SIAs publicly available.

   - **Consult:** Make SIAs open to feedback, considering feedback for future assessments.

   - **Partner:** Collaborate with stakeholders in conducting SIAs, sharing decisions on justifiability.

   - **Empower:** Facilitate stakeholders' independent SIAs, enabling them to decide on justifiability.

2. Have a team discussion reflecting on:

   - The importance of engaging stakeholders, considering the potential impacts posed by this project. Refer to the possible harms and benefits within the stakeholder profiles, considering:

     - Which of the objectives is proportional to the risks that this project may present to salient stakeholders?

   - The strengths and limitations presented by your team positionality —for instance, cases where the identities of team members do not sufficiently reflect or represent salient stakeholders. Refer to the **Hot Air Balloon** section, considering:

     - Which of the objectives could address positionality limitations or "fill the gaps" of missing stakeholder viewpoints through stakeholder participation?

**3.** Reconvene as a team.

**4.** Take a moment to individually consider which engagement objective you believe would equip your impact assessments with a level of engagement which meets assessments of risk and positionality.

**5.** Vote on an engagement objective.

**6.** Having decided on an engagement objective, have a team discussion to further define your engagement objective. Consider:

- How is this objective proportional to the possible impacts that this project may present to stakeholders?

- How does this objective enable your team to address positionality limitations or gaps of missing stakeholder viewpoints pertaining to identity and lived experience?

- What level of influence will stakeholders ideally have on SIAs?

**7.** Your co-facilitator will write your answers in notes, placing them on the **Established Engagement Objective** section. This objective will be recommended to your council as a starting point for assessing the impacts of this project.

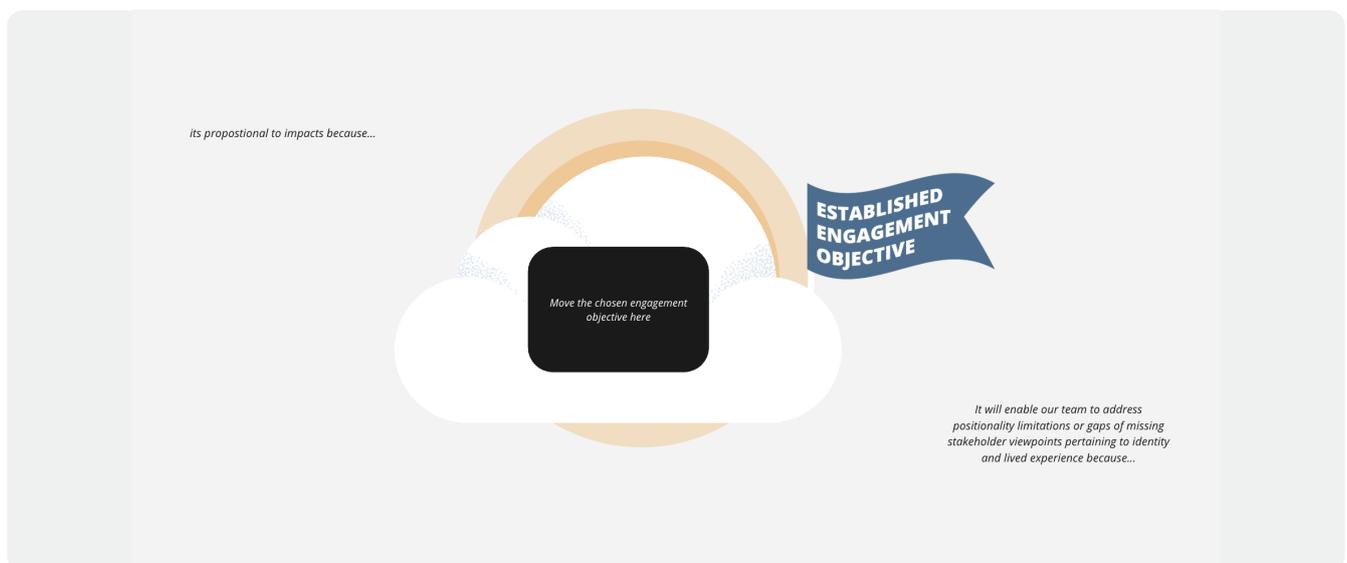

**Established Engagement Objective**

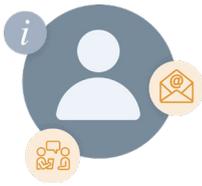

# Establishing an Engagement Objective

**1.** Give the team a moment to read over the activity context and instructions, asking them if they have any questions.

**2.** Split the team into groups.
- Facilitators will lead one group and co-facilitators another.

**3.** In groups, ask the team to look over the different participation goals, using the questions on the Participant Instructions to lead a 15-minute discussion about which goal they believe would be appropriate for this project.

**4.** Reconvene as a team and ask the group to individually vote on an objective.
- **Co-facilitator:** Write team answers in notes, placing them around the Established Engagement Objective on the board.

**5.** Having decided on an engagement objective, lead a 10-minute discussion to further define your engagement objective by considering the questions in the participant instructions.
- **Co-facilitator:** Write team answers in notes, placing them around the **Established Engagement Objective** section.

**6.** Let the team know that you this defined engagement objective will be recommended to the council, their decision on how the project will move forward will be shared in Part Two of this workshop.

# Resources

## Further Explore Modes and Methods of Stakeholder Engagement and Citizen Participation

The Community Planning Toolkit was developed by Community Places. It provides guidance on the issues to consider when planning and designing community engagement.

The Data Justice Lab is a space for research and collaboration at Cardiff University's School of Journalism, Media and Culture (JOMEC). It seeks to advance a research agenda that examines the intricate relationship between datafication and social justice, highlighting the politics and impacts of data-driven processes and big data. They have produced a guidebook for advancing civic participation in algorithmic decision-making.

The Design Justice Network is an international community of people and organisations who are committed to rethinking design processes so that they centre people who are too often marginalised by design. They provide guiding principles and resources for just design practices. See their principles and resources.

*The Innovation in Democracy Programme*, funded and run by the Department for Digital Culture, Media and Sport (DCMS) and the Ministry of Housing, Communities and Local Government (MHCLG), trialled the involvement of citizens in decision-making at local government level through innovative models of deliberative democracy. This programme produced two reports: How to Run a Citizen's Assembly and Innovations in Democracy Programme case studies.

**Involve** is a public participation charity, they develop support and campaign for new ways to involve people in decisions that affect their lives.

- Participatory Methods
- How to Plan a Participatory Process
- Digital Tools for Participation
- Involving Communities in Covid-19 Response and Recovery

*The Local Government Association (LGA)* is a cross-party organisation that works on behalf of councils to ensure local government has a strong, credible voice with national government.

- New Conversations
- New Conversations 2.0
- A councillor's Workbook on Neighborhood and Community Engagement

Participedia is a global crowdsourcing platform for researchers, activists, practitioners, and anyone interested in public participation and democratic innovations. It provides participatory methods, resources, and case studies.

*The Community Foundation of Northern Ireland* has published a directory of civic engagement tools showcasing a variety of engagement methods alongside explanations of how these tools are used and case studies of previous uses.

See also the Stakeholder Analysis and Positionality and Engagement resources and literature contained within the Bibliography and Further Readings section of this workbook.

# **Endnotes**

# Bibliography and Further Readings

## SUM Values

## Engagement

## Stakeholder Analysis and Positionality

# Risk Assessment

HILEG. (2020). Assessment List for Trustworthy Artificial Intelligence (ALTAI) for self-assessment (p. 34). Independent High-Level Expert Group on Artificial Intelligence.

ICO. (2020). Guidance on the AI auditing framework. Information Commissioner's Office. https://ico.org.uk/media/about-the-ico/consultations/2617219/guidance-on-the-ai-auditing-framework-draft-for-consultation.pdf

OECD Framework for the Classification of AI Systems – Public Consultation on Preliminary Findings (p. 58). Organisation for Economic Cooperation and Development. https://oecd.ai/classification

# AI and Urban Planning

Burrowes, K. (2019, June 18). Making Places for Everyone—With Everyone (SSIR). https://ssir.org/articles/entry/making_places_for_everyone_with_everyone

Cabinet Office, & Geospatial Commission. (2021). Planning and Housing Landscape Review — Executive Summary. https://assets.publishing.service.gov.uk/media/61d869588fa8f505893f1c98/Planning_and_Housing_Landscape_Review.pdf

Geospatial Commission. (2019). *Future Technologies Review.* https://www.gov.uk/government/publications/future-technologies-review

Ministry of Housing, Communities & Local Government, & The Rt Hon Esther McVey MP. (2019, November 5). *PropTech dragons form new expert property innovation council.* GOV.UK.https://www.gov.uk/government/news/proptech-dragons-form-new-expert-property-innovation-council

Sideris, N., Bardis, G., Voulodimos, A., Miaoulis, G., & Ghazanfarpour, D. (2019). Using Random Forests on Real-World City Data for Urban Planning in a Visual Semantic Decision Support System. *Sensors (Basel, Switzerland), 19*(10), 2266. https://doi.org/10.3390/s19102266

The Open Data Institute. (2020, August 6). Case study: Unlocking data on brownfield sites. https://theodi.org/article/case-study-unlocking-data-on-brownfield-sites/

# Resources Informing Activities

Boal, A. (2002). *Games for actors and non-actors:* Routledge.

Coimbra, T. C., & Caroli, P. (2020). FunRetrospectives: Activities and Ideas for Making Agile Retrospectives More Engaging. Amazon Digital Services LLC - KDP Print US. https://books.google.co.uk/books?id=1KHMzQEACAAJ

To find out more about the AI Ethics and
Governance in Practice Programme please visit:

turing.ac.uk/ai-ethics-governance

**The
Alan Turing
Institute**